\documentclass[reprint,longbibliography,
 amsmath,amssymb,superscriptaddress]{revtex4-2}
\usepackage{graphicx}% Include figure files
\usepackage{dcolumn}% Align table columns on decimal point
\usepackage{bm}% bold math
\usepackage[colorlinks=true, citecolor=blue, linkcolor=blue, urlcolor=blue]{hyperref}% add hypertext capabilities
\usepackage{xcolor}
\usepackage{amsmath, amssymb, amsfonts}
\usepackage{mathtools}
\usepackage{subfigure}
\usepackage{mathrsfs}
\usepackage{sidecap}
\usepackage{verbatim}

\definecolor{purple1}{rgb}{128,0,128}

\usepackage{bigints}
\newcommand{\bea}{\begin{eqnarray}}
\newcommand{\ea}{\end{eqnarray}}

\definecolor{darkpastelgreen}{rgb}{0.01, 0.75, 0.24}

\def\d{\mathrm{d}}

\begin{document}

\title{%Exact solution of a backreaction model in a quasi-1D condensate
Number-conserving solution for dynamical quantum backreaction\\ in an ultracold bosonic gas} 
\title{%Exact solution of a backreaction model in a quasi-1D condensate
Number-conserving solution for dynamical quantum backreaction\\ in a Bose-Einstein condensate} 
\author{Sang-Shin Baak}
%\email{psn5210@naver.com}
\affiliation{Seoul National University, Department of Physics and Astronomy, Center for Theoretical Physics, Seoul 08826, Korea} 
\author{Caio C. \surname{Holanda Ribeiro}}
%\email{caiocesarribeiro@alumni.usp.br}
\affiliation{Seoul National University, Department of Physics and Astronomy, Center for Theoretical Physics, Seoul 08826, Korea}
\affiliation{Institute of Physics, University of Brasilia, 70919-970 Brasilia, Federal District, Brazil
and International Centre of Physics, University of Brasilia, 70297-400 Brasilia, Federal District, Brazil} 
\author{Uwe R. Fischer}
%\email{uwerf@snu.ac.kr}
\affiliation{Seoul National University, Department of Physics and Astronomy, Center for Theoretical Physics, Seoul 08826, Korea} 

\date\today

\begin{abstract}
We provide a number-conserving approach to the backreaction problem of small quantum fluctuations onto a classical background for the exactly soluble dynamical evolution of a Bose-Einstein condensate,  experimentally realizable in the ultracold gas laboratory.
A force density exerted on the gas particles which is of quantum origin is uniquely identified as the deviation from the classical Eulerian force density. The backreaction equations are then explored for the specific example of a finite size uniform density condensate initially at rest. By assuming that the condensate starts from a non-interacting regime,  and in its ground state, we fix a well-defined initial vacuum condition, which is driven out-of-equilibrium by instantaneously turning on the interactions. The assumption of this initial vacuum accounts for the ambiguity in choosing a vacuum state for interacting condensates, which is  due 
to phase diffusion and the ensuing condensate collapse. 
As a major finding, we reveal that the time evolution of the 
condensate cloud leads to condensate density corrections that cannot in general be disentangled from the quantum depletion in measurements probing the power spectrum of the total density. Furthermore, 
while the condensate is initially at rest, quantum fluctuations give rise to a nontrivial condensate flux, from which we demonstrate that the quantum force density attenuates the classical Eulerian force. Finally, the knowledge of the particle density as a function of time for a condensate at rest
determines, to order $N^0$, where $N$ is the total number of particles, 
the quantum force density, thus offering a viable route for obtaining experimentally accessible quantum backreaction effects. 
\end{abstract}
%We adopt atom number conservation as our guiding principle to render every contribution 
%to the quantum field in the dynamical evolution controllable according to its order, which is given by a %power of the total number of particles $N$. 
%\keywords{Boundary quantum field theory, Quantum vacuum fluctuations, Stochastic processes}

\maketitle

\section{Introduction}

Field quantization in curved spacetimes leads to many intriguing phenomena, some of which have no counterparts in flat spacetimes \cite{Birrell,Wald}.
The most illustrious example, with the largest efficacy in the physics community, is probably the Hawking radiation associated to the formation of event horizons around black holes. 
The emitted quantum particles however backreact onto the spacetime metric containing the horizon, causing it to fluctuate around its classical value.
This {\em backreaction problem} was (partially) addressed 
by Hawking in \cite{Hawking1975} (where backreaction was stated
to be a ``difficult problem") and was entirely ``ignored" in the original announcement 
that black holes radiate \cite{Hawking1}.

%Yet, the measurement of the Hawking temperature remains a daunting problem due to the smallness of the effect in comparison to other sources of heat in the universe. In this sense, in order to study some aspects of this fundamental effect, 
The difficulty (essentially, an impossibility) 
to distinguish the tiny Hawking radiation of real astrophysical black holes from the thermal background which surrounds it in the cosmos has 
%models that mimic the Hawking radiation were devised by the scientific community 
led to the idea of {\it analogues} of Hawking radiation to be implemented in nonrelativistic parent systems \cite{Unruh1981,Visser1998,BLV}, a development which culminated in the recent detection of this analogue Hawking radiation in Bose-Einstein condensates (BECs)  \cite{Steinhauer2019,Steinhauer2021}.

The BEC analogue models are not limited to the simulation of kinematical phenomena 
on curved spacetime such as the Hawking effect, and are lending themselves 
to general studies of backreaction in a system with well established microscopic behavior
 \cite{Fischer2005,Fischer2007}. 
 
%For instance, a black hole analogue can also be used to extract information on how the Hawking-like radiation affects the black hole, i.e., how the radiation backreacts on the analogue model. 
The backreaction problem of small quantum fluctuations onto the condensate in a dilute BEC at first glance might appear to be a straightforwardly soluble one within Bogoliubov theory. A careful inspection however reveals that due to the nontrivial interchange of particles between the condensed and noncondensed (depleted) clouds any proper notion of the separation of a ``classical'' condensate and quantum fluctuations  on top of it can only be developed in a number-conserving theory, with the number of atoms $N$ being fixed at all times. %, henceforth denoted by $N$.
Therefore, the breakdown of particle number-conservation by the standard Bogoliubov expansion %in nonrelativistic condensates 
requires that 
expansions different from the latter should be employed in order to correctly account for backreaction effects, such as 
%the Hartree-Fock-Bogoliubov approach %approximation 
%\cite{Griffin1996} or 
the number-conserving theories as developed, e.g., in \cite{GA1959,Girardeau1998,Gardiner1997,Castin1998,Gardiner2007,Billam2012}. 
In \cite{Fischer2005} the authors considered backreaction by improving the standard Bogoliubov expansion via an expansion in powers of $N$, which is then capable of accounting for the interchange of particles between condensed and depleted clouds. We adopt in our work the same approach, because its predictions for backreaction can be straightforwardly interpreted in terms of 
possible experimental ramifications. 

Another important aspect of backreaction is related to using different field variables (e.g., switching to the quantum Madelung representation instead of using the scalar field operator). Different choices of variables produce conflicting results on the backreaction. This is, for example, illustrated by a comparison of a field-theory-inspired effective action approach to quantum backreaction \cite{Balbinot} with an approach based on  directly measurable  quantities such as full density and current, as performed in \cite{Fischer2005} (also see \cite{Schuetzhold2008}).

Thus, when backreaction is not axiomatically 
defined as for quantum field theory in curved spaces cf.,e.g.,  \cite{Wald1978b,Wald,Birrell}, %corrections to the condensate 
backreaction must be expressed in terms of 
%fundamental quantities, independent of variable choices, which in the case of BEC include 
measurable quantities. % such as particle number and current. 
Following this principle, an unambiguous quantum backreaction definition was established in \cite{Fischer2005} as a force density exerted on the gas particles which leads to a in principle measurable departure from classical Eulerian dynamics. 

In the present paper we explore (and further extend) the quantum backreaction scheme defined in \cite{Fischer2005}, by thoroughly analyzing a concretely soluble dynamical quasi-1D condensate model. We assume that our condensate initially has uniform density; such configurations have been experimentally implemented \cite{Gaunt2013,Navon2021}. The system resides initially in the non-interacting ground state, and  is then suddenly quenched to an interacting regime, with finite coupling constant. 
%Among the applications of uniform density models , for our purposes this assumption is 
The uniform density  assumption leads to a set field equations that can be exactly solved
in the dominant order of condensate corrections.

An important phenomenon occurring in interacting BECs is 
that of condensate phase diffusion \cite{Leggett1991,Imamo,Lewenstein,Castin1998}, 
%,Hongwei2003,Egorov}, 
the %\col{inherent %spontaneous 
%destruction of the condensate due to quantum fluctuations}
 spreading of the condensate phase in a state of fixed particle number $N$. In particular, phase diffusion means that there is no stationary {\em finite size} interacting condensate (note this is distinct from the divergence of phase fluctuations in an {\em infinite} one- and two-dimensional systems \cite{Hohenberg}). 
The absence of a stationary ground state, then, leads to the difficulty of properly specifying the system initial configuration.

By assuming that the condensate is initially non-interacting we provide a clear characterization of the system's {\it initial condition} that, importantly, can be prepared in a controlled way in the laboratory. 
We show in this work that associated to the backreaction problem is a Cauchy problem, that requires the specification of the initial condensate state in order to produce meaningful solutions. For our quenched condensate, starting from a regime where the system is stationary, it is then possible to follow the system evolution in full detail from a well-defined initial state.

%\col{This phase diffusion is of obvious relevance for the dynamical quantum backreaction analysis in a finite size condensate.} 
%\colp{Discuss in our next meeting the connection to experiments: As condensation takes place, it is not possible to assess in which configuration the condensate is when mean field regime is reached.}

From the number-conserving solution for the condensate evolution,  
%in a controlled number-conserving way, 
our model enables the proper interpretation of measurable quantities like density and current of the gas. When the interactions are turned on and quantum fluctuations emerge, we show that it is not possible to discern the quantum depletion from the corrections to the condensed cloud in density measurements, revealing a subtle but important aspect of measuring condensate depletion. Furthermore, we show that even though the condensate is initially at rest and there is no phonon flux, number-conservation leads to a nonvanishing condensed particle flux, representing a bona fide quantum backreaction effect in our system. From the particle current, we then show that the total force density on the gas particles is given by a potential function, which, upon comparison with the Eulerian potential function, is used to show that the quantum force has an attenuating effect over the classical force.   

Our work is organized as follows. We present in section \ref{expansion} the expansion
of the field operator in powers of $N$, and the quantum force definition. In section \ref{quantization}, the condensate model is set up, and the quantum fluctuations 
created by the coupling constant quench are studied. Next, in section \ref{qdepletion}, we study quantum depletion, from which we verify the validity regime of our approach. In section \ref{corrections} we present the major results of our work, the corrections to the condensate resulting from backreaction of quantum fluctuations onto the condensate. We finish our discussion with final remarks in section \ref{remarks}.

\section{Number-conserving formulation of quantum backreaction} %in quasi-1D condensates}
%\label{expansion}

\subsection{Expansion in powers of $N$}
\label{expansion}

Let $\Psi$ describe the bosonic field associated  to the nonrelativistic dilute gas in 1+1 dimensions under study, whose evolution is here taken to be ruled in the s-wave approximation %valid in the dilute-gas approximation
 by the field equation
\begin{equation}
i\partial_t\Psi=\left(-\frac{\partial_x^2}{2m}+U+g|\Psi|^2\right)\Psi,\label{fulleq}
\end{equation}
where we have set $\hbar=1$, and $U$ is the external (trapping) potential. We note that the theory described by \eqref{fulleq} is invariant under global $U(1)$ transformations, which ensures the conservation of the system particle number $N$ according to the Noether theorem. Specifically, the conservation law $\partial_t\rho+\partial_xJ=0$ holds, where $\rho=|\Psi|^2$, $J=(\Psi^{*}\partial_x\Psi-\Psi \partial_x\Psi^*)/2im$ are the system particle and current densities, respectively, and $N=\int \d x\rho$. 
%Here, $\mbox{c. c.}$ stands for complex conjugate.

The number-conserving theory we adopt in this work is obtained by expanding $\Psi$ in powers of $N$ as 
\begin{equation}
\Psi=\phi_{\rm 0}+\chi+\zeta+\mathcal{O}(N^{-3/2}), \label{Eqexpansion}
\end{equation}
with the following set of scaling behaviors \cite{Fischer2005} %hypotheses 
%\col{put citations here where this can be found, in particular $Ng=$const}
\begin{subequations}
\label{ncexpansion}
\begin{align}
&N\rightarrow \infty\quad \mbox{while}\quad Ng=\mbox{constant},\label{cond1}\\
&\phi_{\rm 0}=\mathcal{O}(N^{1/2}), \chi=\mathcal{O}(N^{0}),\ \mbox{and}\ \zeta=\mathcal{O}(N^{-1/2}).\label{cond2}
\end{align}
\end{subequations}
The above set of scalings leads, for the assumed contact interactions,  and at the 
respective orders of $N$, to the hierarchy of coupled equations \eqref{GP}, \eqref{BdG}, and \eqref{GPcorr} below (note that for long-range interactions, in general other scalings  appear  in the expansion, cf.~\cite{Ralf}).  
The first condition \eqref{cond1} is necessary to control the potential $g|\Psi|^2$ in the field equation \eqref{fulleq}, for when $N\rightarrow\infty$ for a fixed coupling $g$, $\Psi\rightarrow 0$ is the only physical (normalizable) solution to the field equation. 
This condition is indeed required for a rigorous derivation of the Gross-Pitaevskii energy functional \cite{LSY2000}, and for complete Bose-Einstein condensation to occur in the limit $N \rightarrow \infty$ \cite{Lieb}.
The second condition \eqref{cond2} ensures that $|\zeta|\ll|\chi|\ll|\phi_{\rm 0}|$ and identifies the different magnitude scales in the system. The field $\chi$ is the fluctuating field and $\zeta$ encapsulates the ``backreaction'' effects from $\chi$ onto the given condensate configuration $\phi_0$. Furthermore, our goal is to work with a number-conserving theory up to order $N^0$, i.e., the densities $\rho$ and $J$ should be expanded also up to order $N^0$. Thus, because both $\rho$ and $J$ are quadratic in $\Psi$, the expansion for $\Psi$ must contain terms up to order $N^{-1/2}$ (i.e., up to the order of $\zeta$). 

By plugging the expansion \eqref{Eqexpansion} into the field equation \eqref{fulleq} 
%and requiring the  holds true at each order of $N$ 
and identifying terms according to their order in $N$ (keeping in mind that $g\propto N^{-1}$), 
we obtain to leading order the Gross-Pitaevskii equation for $\phi_0$ %  This allows us to expand the action in powers of $N$ to obtain
%
%\begin{align}
%S=&\int\d^2x\Bigg\{\phi_{\rm c}^{*}\left[i\partial_t+\frac{\partial_x^2}{2m}-U-\frac{g}{2}|\phi_{\rm c}|^2\right]\phi_{\rm c}\nonumber\\
%&+\chi^{*}\left[i\partial_t+\frac{\partial_x^2}{2m}-U-2g|\phi_{\rm c}|^2\right]\chi\nonumber\\
%&-\frac{g}{2}\left(\phi_{\rm c}^2\chi^{*2}+\mbox{c. c.}\right)\Bigg\}+\mathcal{O}(N^{-1/2}).\label{action}
%\end{align}
%
%
\begin{align}
i\partial_t&\phi_{0}=\left(-\frac{\partial_x^2}{2m}+U+g\rho_0\right)\phi_{0},\label{GP}
\end{align}
$\rho_0=|\phi_0|^2$, in the next order in $N$ 
the Bogoliubov-de Gennes (BdG) equation for $\chi$
\begin{align}
i\partial_t\chi=\left(-\frac{\partial_x^2}{2m}+U+2g\rho_0\right)\chi+g\phi_{\rm 0}^2\chi^{*},\label{BdG}
\end{align}
and finally a  BdG-like equation with $\chi$-dependent  
source terms for the final contribution to the expansion, $\zeta$, 
\begin{align}
i\partial_t&\zeta=\left(-\frac{\partial_x^2}{2m}+U+2g\rho_0\right)\zeta+g\phi_{\rm 0}^2\zeta^{*}+2g|\chi|^2\phi_0\nonumber\\
&\hspace{1cm}+g\chi^2\phi_0^*.\label{GPcorr}
\end{align} 
Alternatively, following \cite{Fischer2005} we can also define the field $\phi_{\rm c}=\phi_0+\zeta$, in such a way that both equations \eqref{GP} and \eqref{GPcorr} are written compactly in terms of the improved Gross-Pitaevskii equation which includes subleading terms 
%containing the anomalous correlator and the depletion density
%
\begin{align}
i\partial_t\phi_{\rm c}=&\left(-\frac{\partial_x^2}{2m}+U+g|\phi_{\rm c}|^2+2g|\chi|^2\right)\phi_{\rm c}+g\chi^2\phi_{\rm c}^{*}.\label{impGP}
\end{align}
It should be stressed, however, that within the expansion in powers of $N$, Eq.~\eqref{impGP} represents {\it only} a compact way of writing Eqs.~\eqref{GP} and \eqref{GPcorr}. 
Indeed, note that factors involving $\chi$ in Eq.~\eqref{impGP} are of order $N^{-1}$, that should be compared with the order $N^{0}$ factors in the remaining terms. Thus, in order to keep consistency of the expansion, the solution $\phi_{\rm c}$ of Eq.~\eqref{impGP} is the sum of a dominant order $N^{1/2}$ term plus a subdominant correction of order $N^{-1/2}$. Using the fields $\phi_{\rm c}$ and $\chi$ to describe the gas dynamics allows us to interpret the field $\zeta$ as modeling corrections to the condensate order parameter $\phi_0$ due to the field $\chi$. 
We will show in the following that this interpretation of the field $\zeta$ facilitates 
the proper formulation of backreaction of quantum fluctuations onto the classical background.%\footnote{Note we depart here from \cite{Fischer2005} which imposed that 
%$\lbracket \zeta \rbracket \propto \ord (N^{-1})$, leading to the necessity of including 
%subleading terms into the GP equation.} 
 
Finally, quantization, within our approach, is achieved by promoting the classical field $\chi$ to the operator-valued distribution $\hat{\chi}$, taken to satisfy the equal-time bosonic commutation relation $[\hat{\chi}(t,x),\hat{\chi}^{\dagger}(t,x')]=\delta(x-x')$ 
\footnote{Note that it is consistent, to leading Bogoliubov order, with 
the expansion \eqref{expansion} and the resulting evolution equations \eqref{GP},\eqref{BdG}, and \eqref{GPcorr}, to quantize the field $\chi$ only (and not also $\zeta$), cf.~the discussion in \cite{Fischer2005}.}.

In this work, we shall consider only a (quasiparticle) vacuum state for the perturbations, i.e., $\langle\hat{\chi}\rangle\coloneqq 0$. After quantization, the classical current and density become operator-valued distributions as well, and for the vacuum state under consideration one has
\begin{align}
&\rho:=\langle\hat{\rho}\rangle=|\phi_{\rm c}|^2+\langle\hat{\chi}^\dagger\hat{\chi}\rangle+\mathcal{O}(N^{-1/2}),\label{dens1}\\
&J:=\langle\hat{J}\rangle=\frac{1}{m}\mbox{Im}\left[ \phi_{\rm c}^{*}\partial_x\phi_{\rm c}+\langle\hat{\chi}^\dagger\partial_x\hat{\chi}\rangle\right]+\mathcal{O}(N^{-1/2}).\label{dens2}
\end{align}
Furthermore, the field $\phi_{\rm c}$ is now given by the operator-valued version of equation \eqref{GP}, where it appears as a multiple of the identity operator. Thus the equation coincides with its vacuum expectation value, and is given by
\begin{align}
&i\partial_t\phi_{\rm c}=\nonumber\\
&\left[-\frac{\partial_x^2}{2m}+U+g|\phi_{\rm c}|^2+2g\langle\hat{\chi}^\dagger\hat{\chi}\rangle\right]\phi_{\rm c}+g\langle\hat{\chi}^2\rangle\phi_{\rm c}^{*},\label{GPave}
\end{align}
where the normal ordering prescription was taken.

It is instructive to define the averaged contributions to the densities stemming from the quantum fluctuations alone as $\rho_{\chi}:=\langle\hat{\chi}^\dagger\hat{\chi}\rangle$ and $J_{\chi}:=\mbox{Im}\ [\langle\hat{\chi}^\dagger\partial_x\hat{\chi}\rangle]/m$, in such a way that $\rho:=\rho_{\rm c}+\rho_{\chi}+\mathcal{O}(N^{-1/2})$ and $J:= J_{\rm c}+J_{\chi}+\mathcal{O}(N^{-1/2})$. It is then straightforward to show from Eqs.~\eqref{GPave} and \eqref{BdG} that
\begin{subequations}
\label{continuity}
\begin{align}
&\partial_t\rho_{\rm c}+\partial_{x}J_{\rm c}=ig(\phi_{\rm c}^2\langle\hat{\chi}^{\dagger2}\rangle-\phi_{\rm c}^{*2}\langle\hat{\chi}^2\rangle),\label{cont1}\\
&\partial_t\rho_{\chi}+\partial_{x}J_{\chi}=-ig(\phi_{\rm c}^2\langle\hat{\chi}^{\dagger2}\rangle-\phi_{\rm c}^{*2}\langle\hat{\chi}^2\rangle),\label{cont2}
\end{align}
\end{subequations}
thus ensuring that the theory is conserving ($\partial_t\rho+\partial_xJ=0$) 
up to our working ($N^{0}$) order in the densities and currents.

\subsection{Corrections to the condensate background}
\label{correctionscon}

We note that the right-hand side of both Eqs.~\eqref{cont1} and \eqref{cont2} are of order $N^{0}$, which is in accord with the left-hand side (L.H.S.) of Eq.~\eqref{cont2}, but seems to fail for the L.H.S. of Eq.~\eqref{cont1}, which is of order $N$. The reason for it is that both $\rho_{\rm c}$ and $J_{\rm c}$ are determined by the field $\phi_{\rm c}$, and thus they split into dominant $\mathcal{O}(N)$ terms plus $\mathcal{O}(N^0)$ corrections, in such a way that the dominant terms in the L.H.S. of Eq.~\eqref{cont1} cancel exactly as they are built from a solution of the standard GP equation \eqref{GP}. 
Returning to the definitions of $\phi_{\rm c}=\phi_{0}+\zeta$ and $\rho_{\rm c}$, $J_{\rm c}$ from Eqs.~\eqref{dens1} and \eqref{dens2}, we find that  $\rho_{\rm c}=\rho_{0}+\rho_{\zeta}+\mathcal{O}(N^{-1/2})$ and $J_{\rm c}=J_0+J_{\zeta}+\mathcal{O}(N^{-1/2})$, where $J_0=\mbox{Im}[\phi^*_0\partial_x\phi_0]/m$,
\begin{align}
\rho_{\zeta}&=2\mbox{Re}\ [\phi_0^*\zeta],\label{rhozeta1}\\
J_{\zeta}&=\frac{1}{m}\mbox{Im}\ [\phi_0^*\partial_x\zeta+(\partial_x\phi_0)\zeta^*].\label{jzeta}
\end{align}
It thus follows from Eq.~\eqref{GP} that $\partial_t\rho_{\rm c}+\partial_xJ_{\rm c}=\partial_t\rho_{\zeta}+\partial_xJ_{\zeta}=\mathcal{O}(N^0)$, ensuring the consistency of Eq.~\eqref{cont1}.

The ``density'' $\rho_{\zeta}$ and ``current density'' $J_{\zeta}$ are the corrections to the condensate contributions $\rho_0$ and $J_0$. Within the validity domain of Bogoliubov theory  i.e., when $\delta N:=\int dx\rho_{\chi}\ll N=\int dx \rho_0$, the quantum fluctuations modeled by the field $\hat{\chi}$ remain small and independent of the condensate corrections $\zeta$, which are in turn determined by $\rho_{\chi}$ and $\langle\hat{\chi}^2\rangle$ through Eq.~\eqref{GPcorr}. In this regime, the dynamics of the field $\hat{\chi}$ is linear. Furthermore, the very condensate existence {\it in the presence of interactions} ($\delta N/N\ll1$) leads to a nonvanishing quantum depletion $\rho_{\chi}$ and nonvanishing anomalous correlator $\langle\hat{\chi}^2\rangle$, which in turn correct the condensate via the field $\zeta$. 
This is the essence of the backreaction scheme we employ here.

\subsection{The quantum force}
\label{qfsec}

We are now able to enunciate the definition of the backreaction scheme presented in \cite{Fischer2005}. The motivation for it comes from the fact that the gas separation into a condensate part modeled by $\phi_{\rm c}$ and the one-particle quantum excitations $\hat{\chi}$ is an intricate concept, as the number of particles in each of these sectors is not conserved separately during the system development. This follows from Eqs.~\eqref{continuity}, which describe the local conservation law of particles in the condensate and depletion sectors.  

Therefore a consistent definition of the quantum backreaction %in this system 
must be formulated in terms of measurable quantities, which, within our nonrelativistic field theory, include the densities and current densities. By following the discussion in \cite{Fischer2005}, in the absence of quantum fluctuations the classical fluid described by the standard GP equation is ruled by the Euler equation $\partial_tJ=f_{\rm cl}$ and the continuity equation $\partial_t\rho+\partial_xJ=0$, where the classical force density $f_{\rm cl}$ is  
\begin{align}
f_{\rm cl}=-\partial_x(\rho v^2)-\frac{\rho}{m}\partial_x\left(\frac{-\partial^2_x\sqrt{\rho}}{2m\sqrt{\rho}}+U+g\rho\right),\label{classical}
\end{align}
and $v=J/\rho$ is the average fluid velocity. However, when quantum fluctuations are taken into account, we should have $\partial_tJ\neq f_{\rm cl}$, indicating a departure from the classical description induced by quantum effects, which unambiguously defines the {\em quantum force}  $f_{\rm q}:=\partial_tJ-f_{\rm cl}$. It then follows from the various definitions in the above that
\begin{align}
f_{\rm q}=&\partial_tJ_{\chi}-v_0\partial_t\rho_{\chi}+\partial_{x}(v_{\rm 0}J_{\chi}-\rho_{\chi}v_{\rm 0}^2)+J_{\chi}\partial_xv_0\nonumber\\
&-\frac{\rho_0}{2m}\partial_x\left(\frac{g G^{(2)}}{\rho_0}\right)+\frac{\rho_{\chi}}{m}\partial_{x}\left(-\frac{\partial^2_{x}\sqrt{\rho_{\rm 0}}}{2m\sqrt{\rho_{\rm 0}}}+U+g\rho_0\right)\nonumber\\
&-\frac{\rho_{\rm 0}}{4m^2}\partial_x\left[\frac{1}{\sqrt{\rho_{\rm 0}}}\partial_{x}^2\left(\frac{\rho_{\chi}}{\sqrt{\rho_{\rm 0}}}\right)-\frac{\rho_{\chi}}{\rho_{\rm 0}^{3/2}}\partial^2_{x}\sqrt{\rho_{\rm 0}}\right]\label{quantumforce}
\end{align}
%
%\col{
%\begin{align*}
%&\Q=\nabla\cdot(\v\otimes\J_\chi+\J_\chi\otimes\v-\rho_\chi \v\otimes\v)-\frac{\rho_0}{2}\nabla \frac{gG^{(2)}}{\rho_0}\nonumber\\
%&-\v(\partial_t\rho_\chi+\nabla\cdot\J_\chi)+\partial_t\J_\chi-\frac{1}{4}\nabla^3\rho_\chi+\rho_\chi\nabla(U+g\rho_0)
%\end{align*}
%}
holds up to order $N^{0}$. Here, $v_0=J_0/\rho_0$ is the zeroth order condensate velocity, and the (second order) correlation $G^{(2)}$ is defined in terms of the quantum density operator $\hat{\rho}=(\phi_{\rm c}^*+\hat{\chi}^\dagger)(\phi_{\rm c}+\hat{\chi})$ 
%(see \cite{Pavloff2012} for an application of density-density correlations in analogue gravity 
as
\begin{equation}
G^{(2)}=\langle:(\hat{\rho}-\langle\hat{\rho}\rangle)^2:\rangle,
\end{equation}   
where the colons indicate normal ordering, 
and is % has a clear interpretation as 
a contribution to the quantum force density 
originating in the local condensate density fluctuations. 

\subsection{The Cauchy problem for the backreaction analysis}
\label{cauchy}

As a final ingredient for the backreaction analysis under construction, in this subsection we show how the various equations presented in the above should be used to calculate relevant quantities. Specifically, we note that the unique identification of classical and quantum forces $f_{\rm cl},f_{\rm q}$ is not sufficient to determine the evolution of $\rho$ and $J$ completely through the system
\begin{subequations}
\label{backreactioneqs}
\begin{align}
\partial_t\rho+\partial_xJ=0,\\
\partial_tJ=f_{\rm cl}+f_{\rm q}.
\end{align}
\end{subequations}
Given the functional forms of both $f_{\rm cl}$ and $f_{\rm q}$ , the system of equations \eqref{backreactioneqs} has to be supplemented with initial conditions for the system densities $\rho$ and $J$ {\it at some initial time}, i.e., the gas configuration should be completely known initially in order for the backreaction problem to become a well-defined Cauchy problem, with a unique causal solution for $\rho$ and $J$.

Furthermore, another subtle aspect of backreaction analysis in Bose gases is linked to the determination of the system initial configuration. Within our expansion scheme, a specification of $\rho$ and $J$ at a given time is characterized by $\phi_0$, $\zeta$, and {\it a quantum state} for the field operator $\hat{\chi}$ at that instant of time. Note that although this specification is simple from a mathematical point of view, it has a great impact on the applicability of the backreaction analysis to experiments devised to probe quantum effects. Indeed, in order to establish a concrete example, let us consider the problem of probing quantum depletion of a condensate through density measurements using for instance the technique of \cite{Hadzibabic2017}. According to the definitions of subsections \ref{expansion} and \ref{correctionscon}, the system total density up to order $N^0$ reads $\rho=\rho_0+\rho_\chi+\rho_\zeta+\mathcal{O}(N^{-1/2})$, and measuring quantum depletion amounts to determining $\rho_\chi$ alone among the various density contributions. In any number-conserving analysis, $\rho_{\chi}$ {\it can only be directly measured in experiments where it can be distinguished from} $\rho_\zeta$. We shall return to this matter when we present our 
approach to backreaction in more detail below. 
%exact backreaction model.

Also related to the notion of initial conditions is the phenomenon of condensate phase diffusion \cite{Castin1998} when $g\neq0$, the spontaneous destruction of the condensate off-diagonal long-range order. For our purposes, this phenomenon means that in the absence of coherent sources no finite size stationary condensate can exist. 
%\col{Here, we should again clearly state that the phase diffusion has nothing to with 
%the divergence of phase fluctuations in 1D and 2D and stems from zero modes in any dimension}
In particular, no finite size condensate with $g\neq0$ and $\partial_t\rho_\chi\coloneqq 0$ is physically possible, even in the case where the field equations are stationary. Within the backreaction language, phase diffusion implies that there is no stationary instantaneous vacuum state that can be used as a quantum state for the initial system configuration, what contributes further to the difficulty in assessing condensate configurations at the laboratory. In the next sections we thoroughly explore a backreaction model with a well-defined initial state that is in principle experimentally accessible.

\section{Quantum fluctuations in a uniform density condensate}
\label{quantization}

%In this section we present the condensate configuration we use for our backreaction analysis and perform the quantization of the field $\chi$. 

As discussed in subsection \ref{cauchy}, within our number-conserving theory, a condensate configuration is specified by the field $\phi_{\rm c}$, or equivalently, by $\phi_0$ and $\zeta$. %, with $\zeta$ depending on $\phi_0$, $\rho_{\chi}$, $\langle\chi^2\rangle$, and appropriate initial conditions to fix an unique solution for $\zeta$. 
We consider a uniform density condensate at rest inside a box trap of size $\ell$ and at zero temperature such that $\rho_0=|\phi_0|^2$ is its density. The condensate is assumed to be in its ground state in a non-interacting regime for $t<0$, i.e., $g\sim0$ (see e.~g.~ \cite{Chen2022} for experimental realizations of condensates in this regime). It thus follows from the absence of interactions that $\rho_\chi=\langle\hat{\chi}^2\rangle\coloneqq0$ ($t<0$) for the system vacuum state, and, accordingly, $\zeta\coloneqq0$ is a trivial solution to the backreaction equations which we take as the initial condensate configuration. In order to activate the quantum fluctuations and thus the backreaction, we assume that at $t\geq0$ the atom interactions are turned on ($g=g_0>0$), while the background condensate order parameter $\phi_0$ remains unchanged. In this section, we study in details the field $\phi_0$ and the quantization of $\chi$ for the resulting condensate configuration. %Equivalently, the condition $\zeta=0$ for $t<0$ translates to $\rho_{\zeta}, J_{\zeta}=0$ for $t<0$ as our initial conditions. \col{In Sec.~\ref{corrections} when we calculate the condensate corrections we show how these conditions enter the backreaction analysis.} 

\subsection{The background condensate}

Condensates with essentially uniform density profile can be prepared using current technology  \cite{Gaunt2013}, 
and we assume for our purposes that the condensate under consideration is well approximated to be uniform. 
This might be realizable ever more accurately experimentally with new trapping techniques being developed (see \cite{Navon2021} for an up-to-date review). A multitude of interesting applications of uniform density models has been described in the latter reference, and we cite here in addition the recent application in the context of analogue gravity 
and Hawking radiation \cite{Curtis}. 

The major benefit we gain by assuming a uniform condensate density is the exact solvability of Eqs.~\eqref{BdG} and \eqref{GPcorr}. Furthermore, we also assume that $\phi_0$ has a simple harmonic time-dependence, i.e., $\phi_0=\exp(-i\mu t)\sqrt{\rho_0}$, where $\mu$ is the chemical potential, and $\rho_0$ is constant. This configuration is a solution of Eq.~\eqref{GP} for the external potential 
\begin{equation}
U=\mu+\frac{1}{2m}\partial_x[\delta(x-\ell/2)-\delta(x+\ell/2)].
\end{equation}
By plugging this external potential back into Eqs.~\eqref{fulleq}, \eqref{GP}, \eqref{BdG}, and \eqref{GPcorr}, and integrating them around $\pm\ell/2$, we obtain that $\partial_x\Psi|_{x=\pm\ell/2}=0$ and similarly for $\phi_0, \chi$, and $\zeta$, i.e., this external potential translates to {\em Neumann boundary conditions} at the condensate walls. In what follows, we shall omit the delta derivatives for the sake of notational convenience.

In order to activate the quantum fluctuations, we assume that at $t\geq0$ the atom interactions are instantaneously turned on ($g=g_0>0$), keeping $\rho_0$ constant. 
This is achieved by a time-dependent external potential given by
\begin{equation}
U=\mu-g\rho_0,
\end{equation}
for $-\ell/2<x<\ell/2$, and this represents all the input necessary to study backreaction in this system, as we shall show.

Once the order $N^{1/2}$ field $\phi_0$ is determined, the condensate quantum fluctuations are given by the quantization of the field $\chi$, and because of the uniform density assumption on $\phi_0$, canonical quantization is straightforward. It consists in expanding $\chi$ in a complete set of eigenfunctions and imposing the canonical commutation relations, as we show in what follows. Also, in addition to $\hbar=1$, we use units such that $m=c_0=\sqrt{g_0\rho_0/m}=1$. Choosing these units, spatial coordinates are expressed in terms of the healing length $\xi_0=1/\sqrt{m\rho_0g_0}=1$, and time is expressed in units of $\xi_0^2$. For instance, particle densities indicate the number of particles per healing length unit.

\subsection{Canonical quantization of $\chi$}

Let $\chi=\exp(-i\mu t )\psi$, where from Eq.~\eqref{BdG} $\psi$ is solution of
\begin{equation}
i\partial_t\psi=-\frac{1}{2}\partial_x^2\psi+\frac{g}{g_0}(\psi+\psi^*),\label{BdGred}
\end{equation}
and we observe the boundary conditions $\partial_x\psi|_{x=\pm\ell/2}=0$. Equation \eqref{BdGred} can be cast in a spinorial form by defining the Nambu spinor $\Phi=(\psi,\psi^*)^{\rm t}$, where ``$\rm t$'' stands for transpose. Thus, the field equation \eqref{BdGred} implies
\begin{equation}
i\sigma_3\partial_t\Phi=\left(-\frac{\partial_x^2}{2}+\frac{g}{g_{0}}\sigma_4\right)\Phi,\label{bogo}
\end{equation}
where $\sigma_i$, $i=1,2,3$ denote the usual Pauli matrices, and $\sigma_4=1+\sigma_1$. Equivalence between the two representations is recovered by requiring that the spinor $\Phi$ satisfies the reflection property: $\Phi=\sigma_1\Phi^*$, and upon quantization the commutation relations in terms of $\hat{\Phi}$ read
\begin{align}
&[\hat{\Phi}_{a}(t,x),\hat{\Phi}^{\dagger}_{b}(t,x')]=\sigma_{3,ab}\delta(x-x'). \label{ccr}
\end{align}
Moreover, the Neumann boundary conditions (BC) for $\psi$ imply that the field $\Phi$ is subjected to the same conditions:
\begin{align}
&\partial_x\Phi|_{x=\pm\ell/2}=0.\label{bc1}
%&\partial_{y}\Phi|_{y=\pm\ell_{\bot}/2}=\partial_{z}\Phi|_{z=\pm\ell_{\bot}/2}=0.\label{bc2}
\end{align}
These boundary conditions together with Eq.~\eqref{bogo} imply that if $\Phi$ and $\Phi'$ are two distinct solutions of the latter equation, then
\begin{equation}
\langle\Phi,\Phi'\rangle=\int\d x\Phi^{\dagger}(t,x)\sigma_3\Phi'(t,x)\label{scalar}
\end{equation}
is a conserved (in time) quantity, which will be used as a scalar product on the space of classical solutions. Also, as the field modes have compact support, they have finite norms, which can be taken in general as
\begin{equation}
\langle\Phi,\Phi\rangle=\pm1.\label{norm}
\end{equation}
We stress that even though Eq.~\eqref{bogo} may admit nonzero solutions with vanishing norm, we can {\it always} find an orthonormal basis as in Eq.~\eqref{norm}. The plus and minus signs in Eq.~\eqref{norm} correspond to positive and negative norm modes, and we recall that for each solution $\Phi$ of Eq.~\eqref{bogo}, $\sigma_1\Phi^*$ is also a solution of opposite norm sign.
Thus there exists a one-to-one correspondence between positive and negative norm modes, which allows us to index the positive norm solutions as $\Phi_n$, $n=0,1,2,\ldots$. With this, we can write the most general classical solution of Eq.~\eqref{bogo} as
\begin{equation}
\Phi(t,x)=\sum_{n=0}^{\infty}\left[a_{n}\Phi_{n}(t,x)+b^{*}_{n}\sigma_1\Phi_{n}^*(t,x)\right],\label{qexpansion}
\end{equation}
and in view of the reflection property $\Phi=\sigma_1\Phi^*$, it follows that $b_{n}=a_{n}$. Now, canonical quantization is defined by the promotion of $\Phi$ to the operator-valued distribution $\hat{\Phi}$ subjected to the condition \eqref{ccr}, which corresponds to promoting each $a_{n}=\langle\Phi_{n},\Phi\rangle$ to an operator $\hat{a}_{n}$ satisfying
\begin{equation}
[\hat{a}_{n},\hat{a}^{\dagger}_{n'}]=\delta_{n,n'}.\label{ccr2}
\end{equation}
Concluding, a vacuum state $|0\rangle$ is defined by the kernel condition $\hat{a}_{n}|0\rangle=0$, and from the identification $\Phi_{n}=(u_{n}, v_{n})^{\rm t}$, we have from Eq.~\eqref{qexpansion}
\begin{equation}
\hat{\psi}(t,x)=\sum_{n=0}^{\infty}\left[\hat{a}_{n}u_{n}(t,x)+\hat{a}^{\dagger}_{n}v_{n}^*(t,x)\right].\label{qfield}
\end{equation}
Therefore, in order to conclude the quantization procedure we need to find the set $\{\Phi_n\}_n$ of positive field modes.   

\subsection{Quantum field in the non-interacting regime}

The positive norm field modes $\Phi_n (t,x)$, $n=0,1,2,\ldots$ at all times can be found by solving the field equation separately at $t<0$ and $t>0$, where the system dynamics is stationary. Let us focus on the non-interacting regime first. In this case, the field equation reads
\begin{equation}
i\sigma_3\partial_t\Phi=-\frac{\partial_x^2}{2}\Phi.\label{bogonon}
\end{equation}
The solutions of Eq.~\eqref{bogonon} can be found as follows. As we are in a stationary regime, solutions of the form $\Phi(t,x)=\exp(-i\omega t)\Phi_{\omega}(x)$ exist for $\omega\geq0$, such that
\begin{equation}
\omega\sigma_3\Phi_{\omega}=-\frac{\partial_x^2}{2}\Phi_{\omega}.\label{bogonon2}
\end{equation}
Now because Eq.~\eqref{bogonon2} for $-\ell/2<x<\ell/2$ is independent of $x$, we can find $\Phi_{\omega}(x)=\exp(i k x)\Phi_{\omega,k}$ with constant $\Phi_{\omega,k}$. This is possible only if
\begin{equation}
\omega=\pm \frac{k^2}{2},
\end{equation} 
which has 4 distinct solutions: $k_1=\sqrt{2\omega}=-k_2=-i k_3=ik_4$, and $\Phi_{\omega,k_1}=\Phi_{\omega,k_2}=(1,0)^{\rm t}$, $\Phi_{\omega,k_3}=\Phi_{\omega,k_4}=(0,1)^{\rm t}$. Thus a general solution of Eq.~\eqref{bogonon2} must have the form
\begin{equation}
\Phi_{\omega}(x)=e^{i k_1 x}\Phi_{\omega,k_1}+\sum_{i=2,3,4}S_{k_{i}}e^{ik_ix}\Phi_{\omega,k_i}.
\end{equation}
By imposing Neumann BC at $x=\pm\ell/2$ we obtain that $S_{k_3}=S_{k_4}=0$, $k_1\coloneqq k_n=n\pi/\ell$, $n=0,1,2,3,\ldots$, $S_{k_2}=(-1)^n$, and $\omega\coloneqq\Omega_n=n^2\pi^2/2l^2$, or
\begin{equation}
\Phi_n(t,x)=\frac{e^{-i\Omega_n t}\left[e^{ik_n x}+(-1)^ne^{-ik_n x}\right]}{\sqrt{2\ell(1+\delta_{0,n})}}(1,0)^{\rm t},\label{modenon}
\end{equation}
$n=0,1,2,3,\ldots$ are the positive norm modes. The normalization constant is added to ensure that $\langle\Phi_n,\Phi_{n'}\rangle=\delta_{n,n'}$. Accordingly, the quantum field in the non-interacting regime assumes the form
\begin{equation}
\hat{\Phi}(t,x)=\sum_{n=0}^{\infty}\left[\hat{a}_{n}\Phi_{n}(t,x)+\hat{a}^\dagger_{n}\sigma_1\Phi^*_{n}(t,x)\right].\label{qf}
\end{equation}
We show in Appendix \ref{completeness} that the for $t<0$ the commutation relation \eqref{ccr} for the expansion \eqref{qf} is verified, which amounts to say that the set of field modes just built is indeed complete.

\subsection{Field modes in the interacting regime}
\label{fieldmodes}

In this subsection we extend the set $\{\Phi_n\}_n$ of positive norm mode functions to the interacting regime ($t>0$). In order to obtain such an extension, we can proceed by solving the field equation for a complete set of mode functions in the interacting regime $(t>0)$ and expand each $\Phi_n$ in terms of these functions. The field equation now reads
\begin{equation}
i\sigma_3\partial_t\Phi=\left(-\frac{\partial_x^2}{2}+\sigma_4\right)\Phi.\label{bogoint}
\end{equation}
In contrast to the non-interacting regime, we call attention to the ``zero norm modes'' caveat of Eq.~\eqref{bogoint}. Note that 
\begin{equation}
\Phi=\Pi_0:=(1,-1)^{\rm t}\label{Pi0}
\end{equation}
 is a time-independent solution of Eq.~\eqref{bogoint}, and because $\Pi_0^\dagger\sigma_3\Pi_0=0$, this nonzero solution has zero norm. Moreover, this is the only time-independent solution of Eq.~\eqref{bogoint}, and clearly $\sigma_1\Pi_0^*=-\Pi_0$, which means that we cannot use the reflection property to build a second linearly independent (LI) solution. A procedure to find another LI solution was presented in \cite{Castin1998}, and in our case it is enough to see that
\begin{equation}
\tilde{\Pi}_{0}=\frac{1}{2}(1,1)^{\rm t}-it \Pi_0\label{pdmode}
\end{equation} 
is also an admissible field mode with zero norm. 
We note that $\tilde{\Pi}_{0}$ is not an eigenfunction of the time translation generator $i\partial_t$: {$i\partial_t\tilde{\Pi}_{0}=\Pi_{0}$, which is the mathematical expression for the breakdown of the system time translation symmetry exhibited by the field equations, implying that no interacting condensate free of external coherent sources exists in a steady state. In the language of Bose-Einstein condensates, the field mode $\Pi_{0}$ corresponds to a ``momentum operator,'' whereas $\tilde{\Pi}_0$ corresponds to an ``unbound phase operator,'' which gives rise to the notion of condensate phase diffusion (see for the definition of these operator notions \cite{Lewenstein,Castin1998}).
%: No interacting condensate free of external coherent sources exists in a steady state. Bona fide field modes with positive norm can then be constructed by taking linear combinations of $\Pi_0, \tilde{\Pi}_{0}$ by exploring the property $\langle \Pi_0,\tilde{\Pi}_0\rangle\neq0$.} 
%For our purposes, though, only the LI character of such modes is important, as we need only a complete set, not necessarily orthonormal. \col{why so? below we normalize} 

All the other positive norm mode functions of Eq.~\eqref{bogoint} can be found following the same procedure as before. We find that
\begin{equation}
\Pi_n=\frac{e^{-i\omega_n t}\left[e^{ik_n x}+(-1)^ne^{-ik_n x}\right]}{\sqrt{2\ell[1-(\omega_n-k_n^2/2-1)^2]}}(1,\omega_n-k_n^2/2-1)^{\rm t}, \label{Pin}
\end{equation}
for $n=1,2,3,\ldots$, and $\omega_n=\sqrt{k_n^2(k_n^2/4+1)}$. Again, the normalization constant was chosen as to guarantee $\langle \Pi_{n},\Pi_{n'}\rangle=\delta_{n,n'}$, $n,n'\geq1$. Therefore, the set $\{\Pi_0,\tilde{\Pi}_{0},\Pi_n,\sigma_1\Pi_n^*\}$ is complete, and thus we can write
\begin{equation}
\Phi_n=\alpha_{n,0}\Pi_0+\beta_{n,0}\tilde{\Pi}_0+\sum_{j=1}^{\infty}[\alpha_{n,j}\Pi_j-\beta_{n,j}\sigma_1\Pi_j^*],\label{series}
\end{equation}
for $t>0$. The coefficients $\alpha_{n,j}$ and $\beta_{n,j}$ are then fixed by the field equation \eqref{bogo}: The wave function $\Phi_n$ is a continuous function of $t$. In particular, we have $\Phi_n(0^+,x)=\Phi_n(0^-,x):=\Phi_n^{(-)}$. This condition applied to Eqs.~\eqref{modenon} and \eqref{series} gives rise to a Fourier decomposition for the function $\Phi_n^{(-)}$, implying
\begin{align}
&\alpha_{n,0}=\frac{\langle\tilde{\Pi}_0,\Phi_n^{(-)}\rangle}{\langle\tilde{\Pi}_0,\Pi_0\rangle},\ \ \ \alpha_{n,j}=\langle\Pi_j,\Phi_n^{(-)}\rangle,\\
&\beta_{n,0}=\frac{\langle\Pi_0,\Phi_n^{(-)}\rangle}{\langle\Pi_0,\tilde{\Pi}_0\rangle},\ \ \ \beta_{n,j}=\langle\sigma_1\Pi_j^*,\Phi_n^{(-)}\rangle,
\end{align}
where $j>0$ and the functions $\Pi_n$ are evaluated at $t=0$. By performing the integrals we find
\begin{subequations}
\label{coeffm}
\begin{align}
&\alpha_{n,0}=\frac{\delta_{n,0}}{2\sqrt{\ell}},\ \alpha_{n,j}=\frac{\delta_{n,j}}{\sqrt{1-(\omega_n-k_n^2/2-1)^2}},\\
&\beta_{n,0}=\frac{\delta_{n,0}}{\sqrt{\ell}},\ \beta_{n,j}=\frac{(\omega_n-k_n^2/2-1)(-1)^n\delta_{n,j}}{\sqrt{1-(\omega_n-k_n^2/2-1)^2}},
\end{align}
\end{subequations}
$j>0$. This concludes the determination of the complete set of positive norm mode functions $\{\Phi_n\}_n$, and we show in Appendix \ref{completeness} that the quantum field expansion of Eq.~\eqref{qf} satisfies the commutation relation of Eq.~\eqref{ccr} for all $t$. In the next section we show how this quantization determines the evolution of the noncondensed cloud.

\section{The condensate depletion}
\label{qdepletion}

In this section we focus on the evolution of the depleted cloud as the interactions are turned on. It should be stressed that the quantization developed in the last section and the results presented in this section are exactly the same as the ones obtained from the non-number-conserving Bogoliubov theory. Major differences between the number- and non-number conserving approaches are found when we discuss connections to measurements in section \ref{corrections}.

The depleted cloud is characterized by the quantum depletion $\rho_{\chi}$ and the phonon flux $J_{\chi}$, both defined in terms of $\hat{\chi}$ in the paragraph before Eqs.~\eqref{continuity} as $\rho_\chi=\langle\hat{\chi}^\dagger\hat{\chi}\rangle$, $J_\chi=\mbox{Im}\ [\langle\hat{\chi}^\dagger\partial_x\hat{\chi}\rangle]$. Referring back to the quantum field expansion in Eq.~\eqref{qfield}, and the definition $\hat{\chi}=\exp(-i\mu t)\hat{\psi}$, for our condensate configuration it follows that $J_{\chi}=0$ for all $t$, meaning that the depleted cloud remains at rest with respect to the laboratory frame as long as the predictions of Bogoliubov theory are reliable.

As for the quantum depletion, we find that   $\rho_{\chi}\coloneqq 0$ for $t<0$ in the non-interacting regime by definition of the latter and
\begin{widetext}
\begin{equation}
\rho_{\chi}=\frac{t^2}{\ell}+\frac{1}{2\ell}\sum_{n=1}^{\infty}\frac{(-1)^n}{\omega_{n}^2}\left[(-1)^n+\cos (2k_nx)\right]\left[1-\cos(2\omega_n t)\right],\label{depletion}
\end{equation}  
\end{widetext}
for $t\geq0$ (interacting regime). Inspection of Eq.~\eqref{depletion} reveals that the first contribution to the system depletion, namely, $t^2/\ell$, comes from the zero mode in Eq.~\eqref{pdmode}. It has a simple interpretation %represents the contribution to depletion coming from the 
in terms of condensate phase diffusion \cite{Castin1998}: As the condensate phase degrades, particles leave the condensed cloud. Moreover, we see that if $\ell\rightarrow\infty$ with $\rho_0$ kept constant, this contribution goes to zero. However, the second contribution diverges then, as {\it no infinite quasi-1D condensate exists} \cite{Hohenberg}. 

In Fig.~\ref{figmain1} we plot some depletion profiles for two system sizes: $\ell=40$ and $\ell=100$,
\begin{figure}[b]
\includegraphics[scale=0.48]{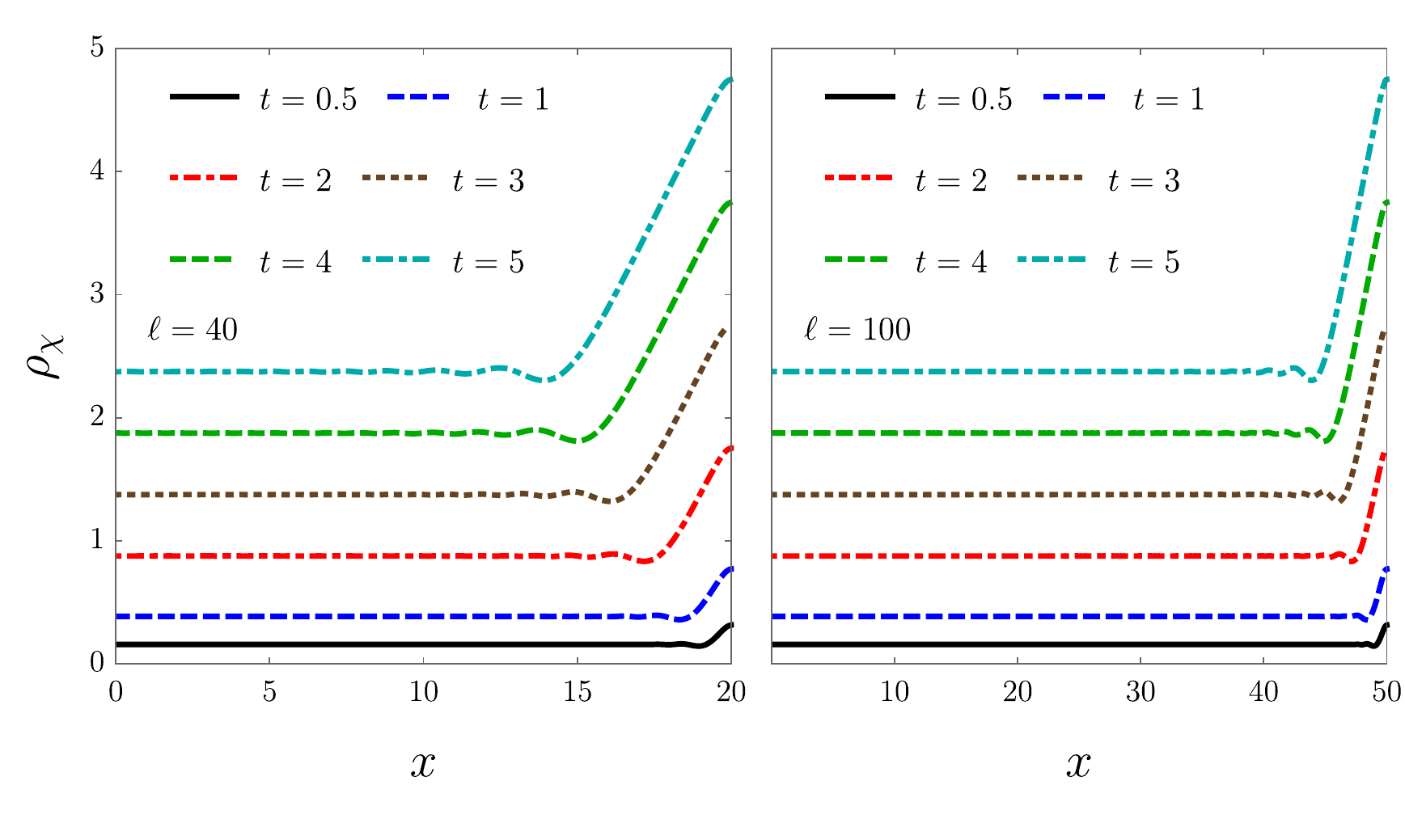}
\caption{Left panel: Evolution of the condensate depletion for a condensate of size $\ell=40$. The curves are plotted for $x\geq0$ only, using that $\rho_{\chi}$ is an even function of $x$ [see Eq.~\eqref{depletion}]. As time passes, we observe an overall depletion increase, initially more pronounced at the condensate wall at $x=\ell/2$.  Right panel: Depletion profile evolution for a system of size $\ell=100$. We note that the bulk depletion increase is insensitive to the existence of the condensate walls for the time periods considered in the simulation.}
\label{figmain1}
\end{figure}  
from which we observe that when the interactions are turned on, depletion increases from zero following a pattern such that far from the condensate walls the system is insensitive to the existence of the (von Neumann)  boundary conditions, as can be inferred from the comparison between the plots in the left and right panels of Fig.~\ref{figmain1}. Furthermore, as times passes, depletion increases faster closer to the condensate walls for both system sizes, giving rise to a wave with growing amplitude in the depleted cloud, that propagates towards the condensate bulk. Moreover, returning to Fig.~\ref{figmain1} left and right panels, in addition to the observed depletion insensitivity on the system size far from the walls, it also follows that the depletion profiles for both sizes are similar close to the walls, as shown in Fig.~\ref{figmain2} left panel for several system sizes at fixed time $t=5$. 

% \col{This property, together with the phase diffusion term of Eq.~\eqref{depletion}, which shows that for smaller systems sizes phase diffusion contributes more to depletion, reveals that it is not possible to disentangle the various depletion sources in the system in terms of fundamental phenomena like phase diffusion.} 
%
\begin{figure}[b]
\includegraphics[scale=0.45]{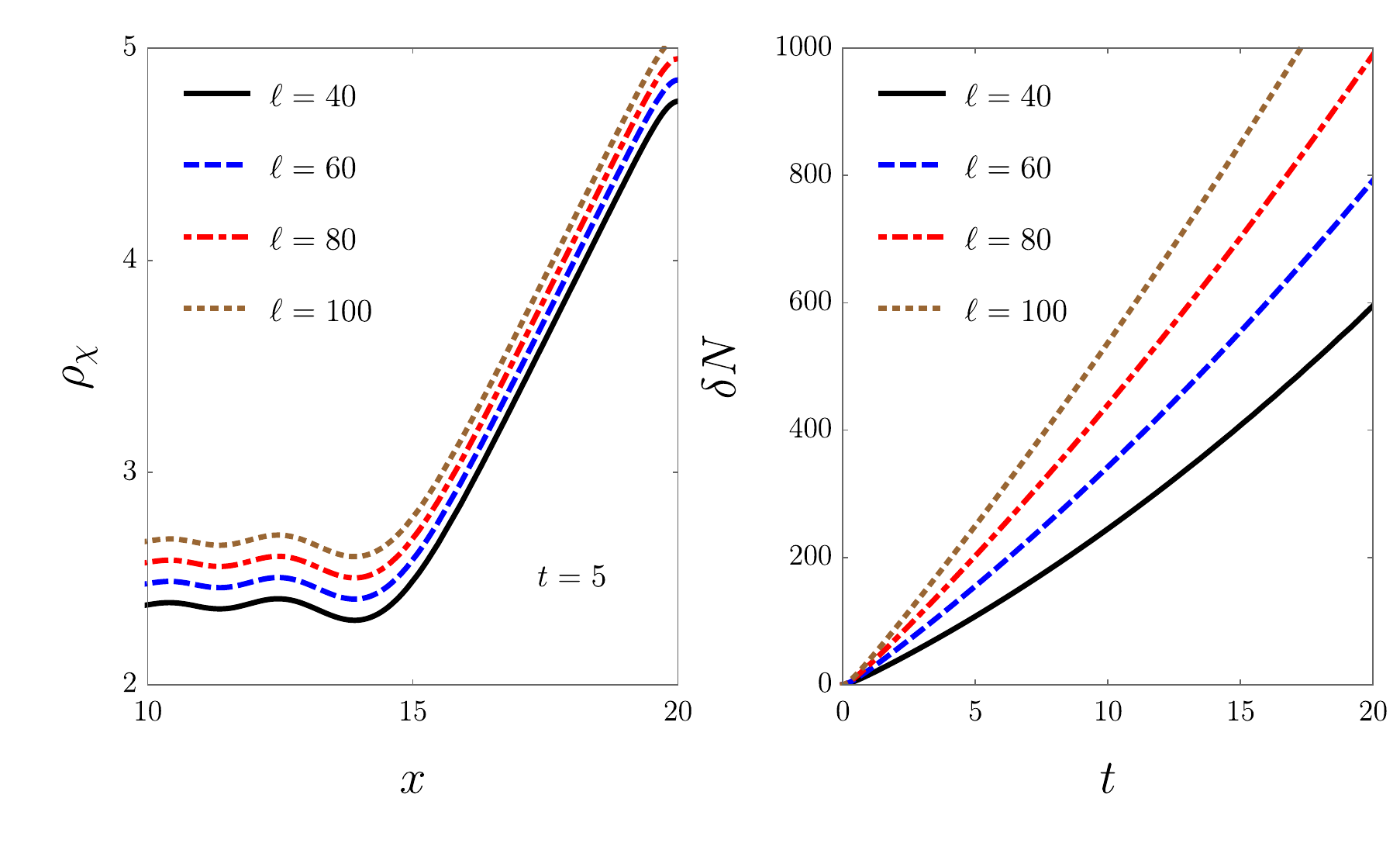}
\caption{Left panel: Depletion near-boundary behavior for several system sizes at $t=5$. The profiles corresponding to larger condensates are translated to the left, and slightly shifted as to allow comparison with the smaller condensate profile. Right panel: Total number of particles in the depleted cloud as function of time for several condensate sizes. Larger condensates correspond to faster growth of $\delta N$ for fixed $\rho_0$.}
\label{figmain2}
\end{figure}  

From our observation that far from the system walls the depletion growth rate is rather insensitive to the system size (cf. both panels in Fig.~\ref{figmain1}), it follows that the growth time scale depends only on the chemical potential $\rho_0g_0$ (remember $t$ is in units of $1/\rho_0g_0$). Moreover, we recall that the quantum depletion of an infinitely extended homogeneous 3D condensate in its ground state is proportional to $\sqrt{\rho_{\rm 3D}g_{\rm 3D}^3}$ \cite{Yang1957,Hadzibabic2017}. Yet, no analogous formula exists for the 1D condensate due to the infrared depletion divergence \cite{Hohenberg}, whereas for (inhomogeneous) finite size condensates the depletion dependence on $\rho_0g_0$ is necessarily model-dependent. %, and must take into account the condensate phase diffusion
In our model, depletion is time dependent, and we can obtain information regarding the depletion growth rate dependence on the condensate chemical potential 
$\rho_0g_0$ far from the condensate walls for a fixed time duration. Indeed, the left panel of Fig.~\ref{figmain1} suggests that within the interval $1\leq t\leq5$ the depletion growth far from the condensate walls ($x\sim0$) is fairly linear with $t$. Within our conventions, $t$ is expressed in units of $\xi_0^2=1/\rho_0g_0$, and thus to return to dimensionful time we must put %change explicit time dependencies as 
$t\rightarrow  \rho_0g_0t$. Hence for a fixed, parameter-independent time period, the quantum depletion in our model, after the interactions are turned on, is approximately linear in $\rho_0g_0$, which should be compared with the 3D counterpart which is $\propto\sqrt{\rho_{\rm 3D}g_{\rm 3D}^3}$.
 
Before solving the backreaction problem in the next section, we note an aspect of the quantum depletion which is of basic importance for our analysis. Namely the validity of the Bogoliubov expansion, and consequently, of the number-conserving expansion presented before Eqs.~\eqref{ncexpansion}, expressed by the condition $\delta N/N\ll1$, where $\delta N$ is the number of depleted particles, which from Eq.~\eqref{depletion} is found to be
\begin{equation}
\delta N=t^2+\frac{1}{2}\sum_{n=1}^{\infty}\frac{1}{\omega_{n}^2}\left[1-\cos(2\omega_n t)\right].\label{ndepatom}
\end{equation}
We plot in Fig.~\ref{figmain2} right panel $\delta N$ as function of time for several system sizes. We observe that for larger sizes $\delta N$ grows faster, which is expected as larger systems have more particles for a fixed $\rho_0$. Furthermore, for a system with size $\ell=40$, we notice that for $t=20$, $\delta N\sim600$. By assuming a condensate with $N=5000$ particles, we find $\delta N/N=0.12$. For definiteness, in this work we assume $\delta N/N\lesssim0.1$ as the validity regime for the field expansion in powers of $N$. Thus, for $\delta N/N=0.12$ the results that follow from the expansion might accordingly not be reliable. We observe that for $N=5000$, the results of Fig.~\ref{figmain1} are within the expansion validity.

%
%\begin{figure}[t]
%\includegraphics[scale=0.4]{figmain3.pdf}
%\caption{}
%\label{figmain3}
%\end{figure}  
% 

\section{Quantum backreaction}
\label{corrections}

In this section we present the major results of our analysis, namely, the (up to the relevant order in $N$) exact solutions for the condensate corrections $\rho_{\zeta}$ and $J_{\zeta}$ in the interacting regime (i.~e.~for times $t>0$). These quantities are determined by the coupled system of equations $(J_0=J_\chi=0)$
\begin{subequations}
\label{backreaction}
\begin{align}
&\partial_t\rho_{\zeta}+\partial_{x}J_{\zeta}=ig(\phi_{0}^2\langle\hat{\chi}^{\dagger2}\rangle-\phi_{0}^{*2}\langle\hat{\chi}^2\rangle),\\
&\partial_tJ_{\zeta}=f_{\rm cl}+f_{\rm q},
\end{align}
\end{subequations}
subject to the initial conditions $\rho_{\zeta},J_{\zeta}=0$ at $t=0$. %Also, recall that $f_{\rm cl}$ depends linearly on $\rho_{\zeta}$ through Eq.~\eqref{classical}. 
This initial value problem then gives rise to the unique solution to the problem of how the condensate evolves during the unavoidable depleted cloud formation. 

The system \eqref{backreaction} plus initial conditions can be solved numerically with the aid of the correlations $\rho_{\chi}, \langle\hat{\chi}^2\rangle$ calculated from the quantum field $\hat{\chi}$. Notwithstanding, for the particular condensate model adopted in our work, the exact solution for the backreaction problem can also be constructed by solving directly for the field $\zeta$ in Eq.~\eqref{GPcorr} instead, from which the $\rho_{\zeta}, J_{\zeta}$ can be determined. We describe in Appendix \ref{cr} the (rather cumbersome) construction of the solution for $\zeta$: $\zeta=0$ for $t<0$ and the exact solution for $t\ge 0$ reads 
\begin{widetext}
\begin{align}
e^{i\mu t }\sqrt{\rho_0}\zeta=-\frac{t^2}{2\ell}-\frac{1}{8\ell}\sum_{n=1}^\infty\frac{(-1)^n}{\omega_n^2}\Bigg\{&2(-1)^n\left[1-\cos2\omega_n t+ik_n^2\left(\frac{\sin 2\omega_nt}{2\omega_n}-t\right)\right]+\cos2k_nx\bigg[\frac{2-k_n^2}{k_n^2+1}+2\cos2\omega_nt\nonumber\\
&-\frac{4i\omega_n}{k_n^2}\sin2\omega_nt-\frac{k_n^2+4}{k_n^2+1}\left(\cos\omega_{2n}t-\frac{i\omega_{2n}}{2k_n^2}\sin\omega_{2n}t\right)\bigg]\Bigg\},\label{zetasol}
\end{align}
\end{widetext}
which contains all the backreaction information within our number-conversing expansion
to the relevant order. We now discuss separately the implications of $\zeta$, as specified  by Eq.~\eqref{zetasol}.

\subsection{The gas density}
\label{gasdensity}

Let us start by studying the gas density $\rho=\rho_0+\rho_\chi+\rho_\zeta+\mathcal{O}(N^{-1/2})$. As the interactions are turned on, $\rho_0$ remains constant, and $\rho_\chi$, which models the evolution of the depleted gas cloud, was studied in the previous section. Now, as a response to the evolution of $\rho_\chi$ dictated by the atom number-conservation, the condensate density $\rho_0+\rho_\zeta$ is corrected by the function $\rho_\zeta$, of the same order ($N^{0}$) as $\rho_\chi$. We have from Eq.~\eqref{rhozeta1} that $\rho_\zeta=2\mbox{Re}\ [\exp(i\mu t)\sqrt{\rho_0}\zeta]$, and thus
\begin{widetext}
\begin{equation}
\rho_\zeta=-\frac{t^2}{\ell}-\frac{1}{4\ell}\sum_{n=1}^{\infty}\frac{(-1)^n}{\omega_n^2}\left\{2(-1)^n[1-\cos(2\omega_n t)]+\cos(2k_n x)\left[\frac{2-k_n^2}{1+k_n^2}+2\cos(2\omega_n t)-\frac{k_n^2+4}{k_n^2+1}\cos(\omega_{2n}t)\right]\right\}.\label{rhozeta2} 
\end{equation}
\end{widetext}
We note first that $\rho_\zeta$ given by Eq.~\eqref{rhozeta2} is not proportional to $\rho_\chi$ and it is such that $\int \d x\rho_\zeta=-\delta N$, where $\delta N$ is the number of depleted particles given by Eq.~\eqref{ndepatom}. Therefore, we have $\int \d x \rho=N$, and we verify that the total number of particles is indeed preserved up to order $N^{0}$.

We present in Fig.~\ref{figmain3} selected plots for $\rho_\zeta$.
\begin{figure}[t]
\includegraphics[scale=0.48]{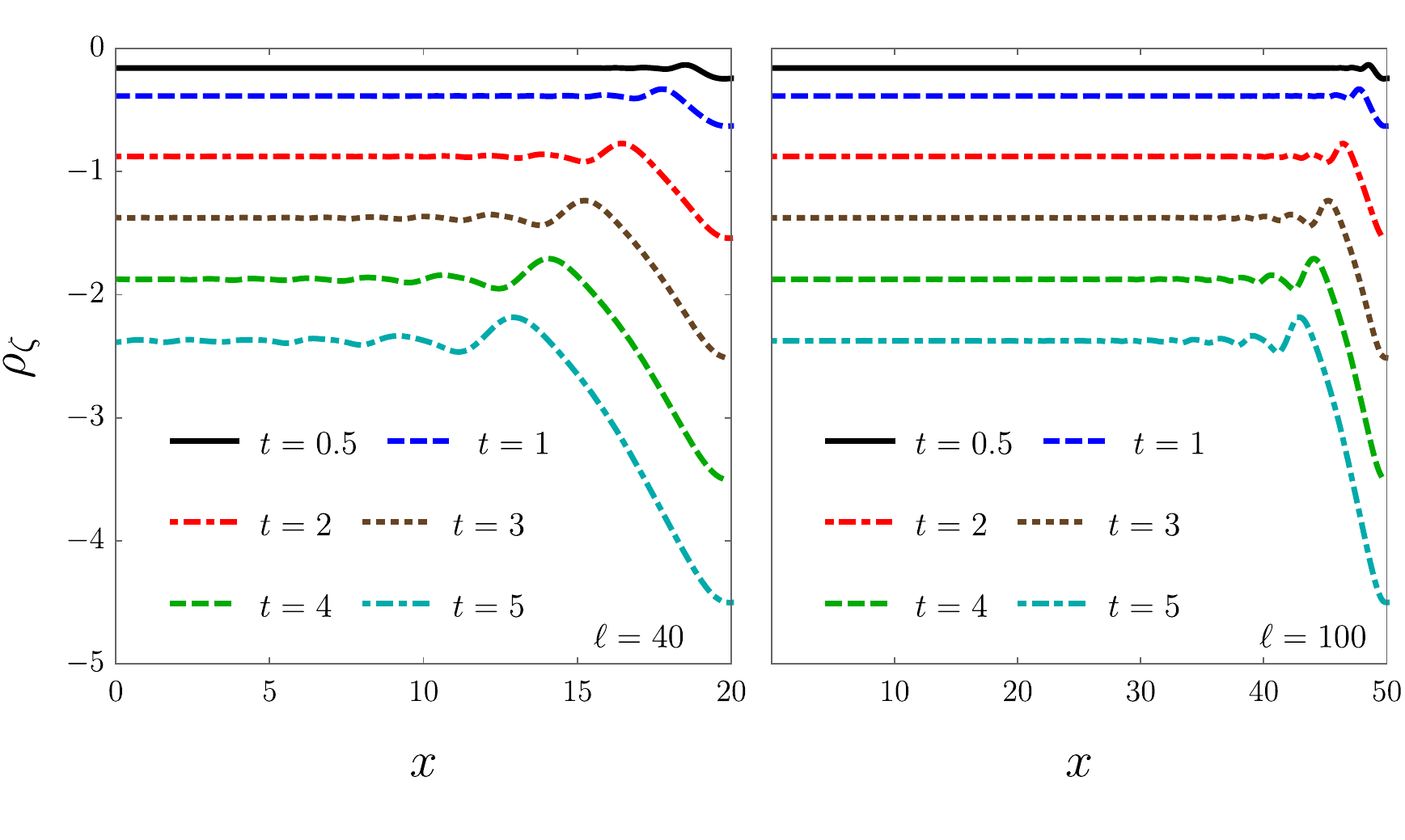}
\caption{Left panel: Evolution of the condensate correction for a condensate of size $\ell=40$. The curves are plotted for $x\geq0$ as $\rho_\zeta$ is an even function of $x$ [see Eq.~\eqref{depletion}]. As time passes, we observe an overall depletion increase, initially more pronounced at the condensate wall at $x=\ell/2$.  Right panel: Depletion profile evolution for a system of size $\ell=100$. We note that the bulk depletion increase is insensitive to the existence of the condensate wall boundary region for the time scales considered in the simulation.}
\label{figmain3}
\end{figure}  
We notice that $\rho_\zeta<0$ for all the simulations of Fig.~\ref{figmain3}, indicating the decrease of the number of condensed atoms when depletion occurs. Furthermore, $\rho_\zeta$ shares some of the properties of $\rho_\chi$ discussed in section \ref{qdepletion}, namely, the density of particles leaving the condensate is fairly insensitive to the system size for the time interval considered in the simulation of Fig.~\ref{figmain3}.

An important implication of our backreaction solution is linked to measurement processes in condensates. Let us consider, for instance, that an experiment is devised to determine quantum depletion in a condensate using the Bragg scattering technique of \cite{Hadzibabic2017}, where the authors explored the fact that for a particular condensate configuration the power spectrum of $\rho_0$ was exponentially suppressed in comparison to the power spectrum of $\rho_\chi$. Thus a full density measurement is in principle capable of separating $\rho_0$ from $\rho_\chi$. However, for a number-conserving analysis such as the one adopted in our model, density measurements can only detect the sum $\rho_{\chi}+\rho_{\zeta}$ (Fig.~\ref{figmain4}) on top of the (in our model) constant background  $\rho_0$. 
\begin{figure}[b]
\includegraphics[scale=0.6]{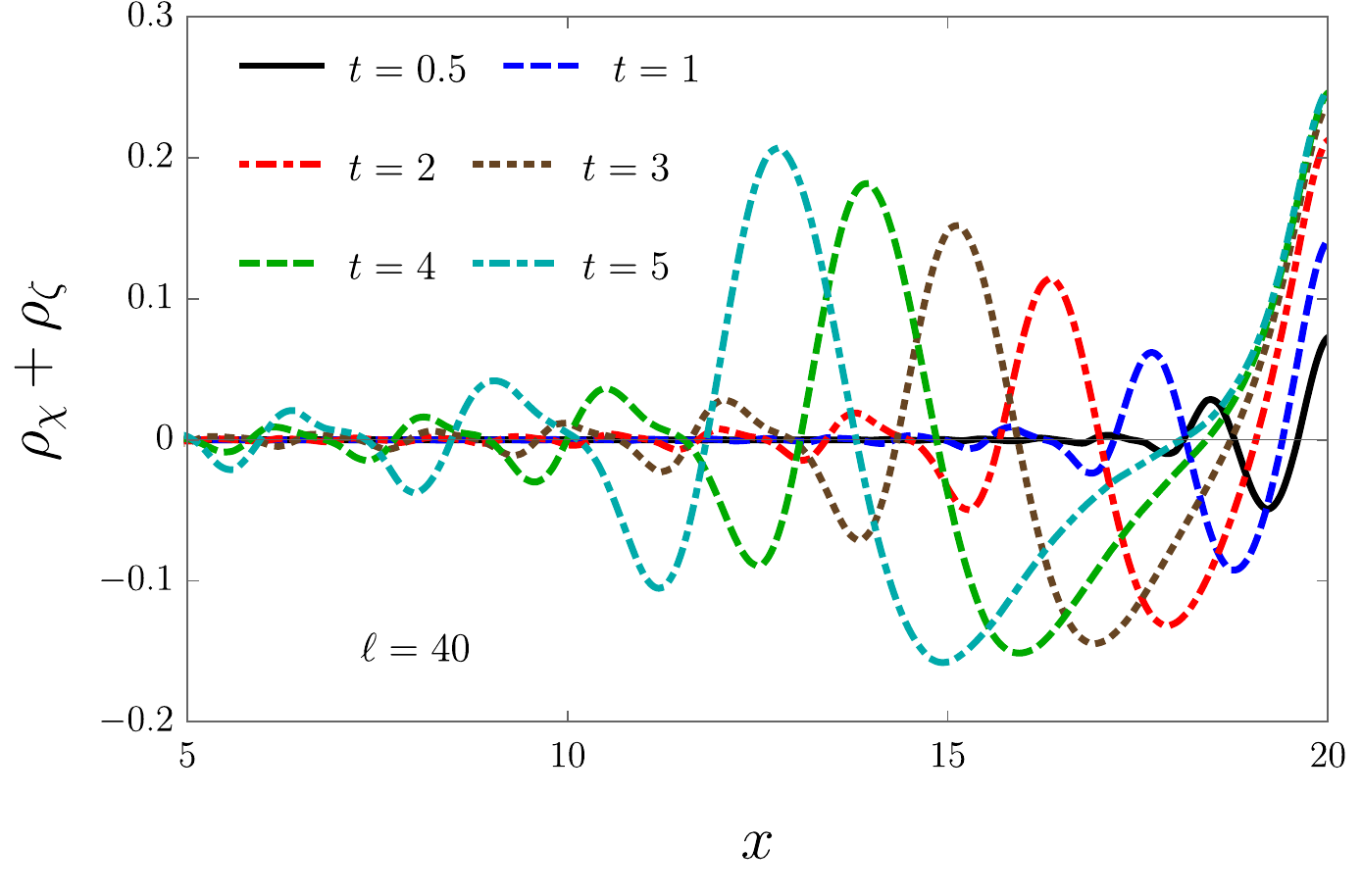}
\caption{Evolution of the gas density on top of the condensate background $\rho_0$ for a system of size $\ell=40$ and at several instants of time. These profiles represent the departure from an uniform density condensate profile as dictated by number-conserving backreation effects.}
\label{figmain4}
\end{figure}  
Therefore, if no feature exists in $\rho_\chi$ distinguishing it from $\rho_\zeta$, 
cf.~Figs.~\ref{figmain1} and \ref{figmain3} representing the buildup of 
$\rho_\chi$ and $\rho_\zeta$, respectively, 
it is not possible to determine $\rho_\chi$ separately by the density power spectrum, or in general, via any measurement relying on an analysis of only the total density. In particular, our analysis also reveals that number-conservation renders some measurement processes in condensates sensitive to how the condensate is prepared. 

To illustrate this further, in Fig.~\ref{figmain4}, we plot $\rho_\chi+\rho_\zeta$ for a condensate of size $\ell=40$ for several instants of time. We observe that as time passes an oscillatory pattern emerges on top of the condensate density, which might be visible in the power spectrum of $\rho_\chi+\rho_\zeta$, and thus can be measured.

\subsection{Induced condensate flow}

In this subsection we discuss perhaps the most intriguing feature coming from the number conservation. As the interactions are turned on, even though there is no flux of depleted particles, the condensate particles undergo a nontrivial flow coming from the backreaction. We recall that the total flux of particles in the system is given by $J=J_0+J_\chi+J_{\zeta}+\mathcal{O}(N^{-1/2})$, and thus for our model, in which $J_0=J_\chi=0$, a condensate flow is modeled by a nonzero $J_\zeta$. That $J_\zeta$ is necessarily nonzero follows directly from the number-conservation of Eqs.~\eqref{continuity} and $\rho_\chi+\rho_\zeta\neq0$: $\partial_xJ_{\zeta}=-\partial_t(\rho_\chi+\rho_\zeta)\neq0$, and thus we must have $J_\zeta\neq0$ during the condensate evolution. The particular form of $J_\zeta$ can be calculated with the aid of Eq.~\eqref{jzeta}, and we find, using \eqref{zetasol},   that $J_\zeta=\mbox{Im}\ [\phi_0^*\partial_x\zeta]$ is given by
\begin{equation}
J_{\zeta}=\frac{-2}{\ell}\sum_{n=1}^\infty\frac{(-1)^n\sin(2k_nx)}{k_n}\left[\frac{\sin(2\omega_nt)}{2\omega_n}-\frac{\sin(\omega_{2n}t)}{\omega_{2n}}\right].\label{jzetaex}
\end{equation}
We present in Fig.~\ref{figmain5} the evolution of the particle flux $J_{\zeta}$ for the condensate of size $\ell=40$ presented in Fig.~\ref{figmain4}.
\begin{figure}[t]
\includegraphics[scale=0.6]{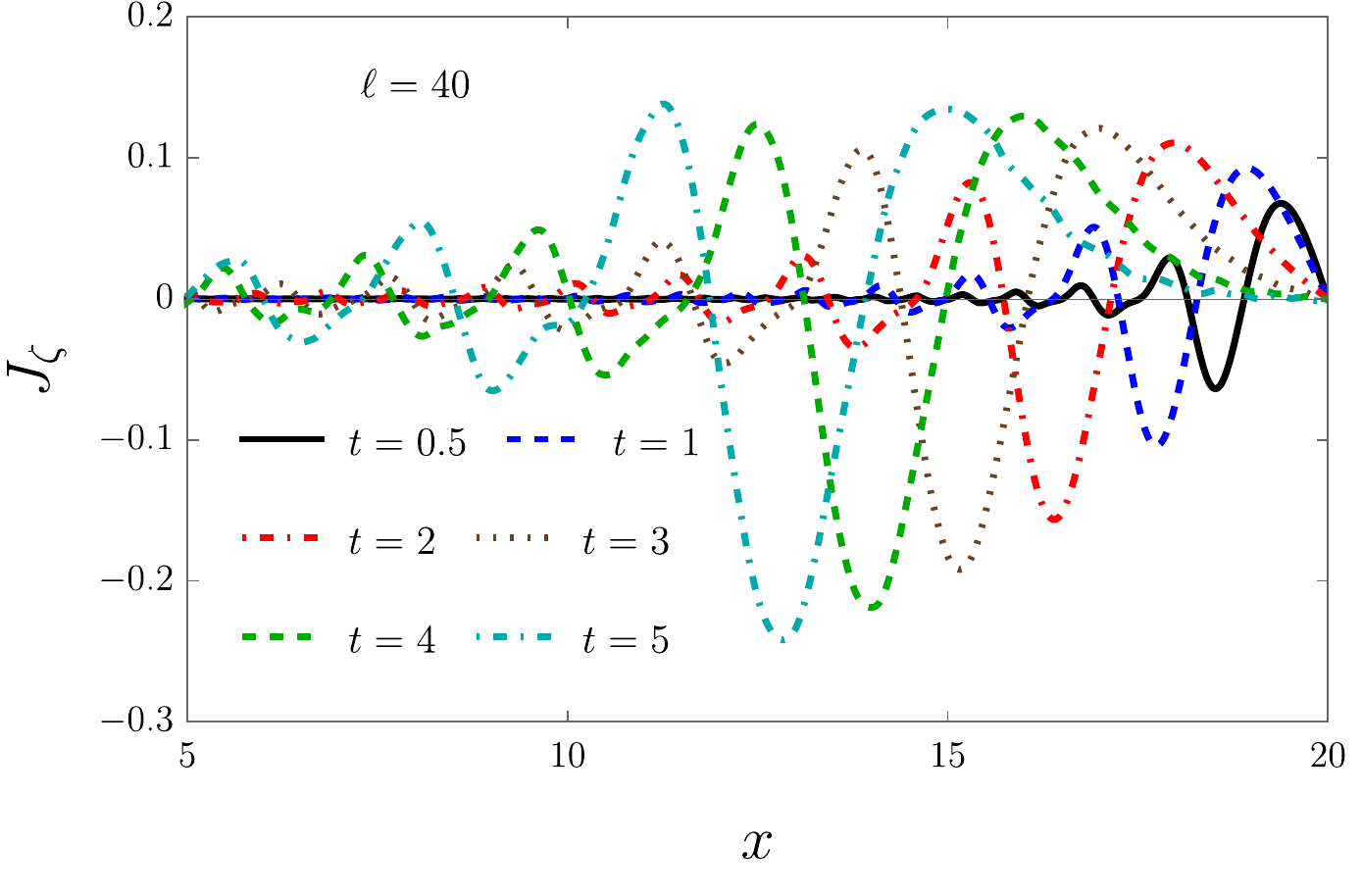}
\caption{Evolution of the condensate flux $J_{\zeta}$ for a condensate of size $\ell=40$ and at several instants of time. The positive plot range $5<x<20$ is motivated by the fact that $J_{\zeta}$ is an odd function in view of Eq.~\eqref{jzetaex}. Note that the flux of particles vanishes at the condensate walls, reflecting the fact that the particles are indeed trapped inside the box.}
\label{figmain5}
\end{figure}  

We note that although Eq.~\eqref{jzetaex} was found from the exact solution of Eq.~\eqref{zetasol}, as discussed in the above, the continuity equation, when $J_\chi=0$, reads   $\partial_xJ_\zeta=-\partial_t(\rho_\chi+\rho_\zeta)$ and can be integrated to find $J_{\zeta}$. Thus, number-conservation implies that we can always find $J_{\zeta}$ in any model where $J_0=J_\chi=0$ and the full density is known, offering a route for determining the particle flux from density measurements.     

\subsection{The quantum potential}

In principle, the only quantized field in our condensate treatment is $\hat{\chi}$ and,  accordingly, all the effects coming from $\hat{\chi}$ have a ``quantum origin.'' That includes $\rho_\zeta$ and $J_{\zeta}$, and justifies the backreaction nomenclature because in the absence of $\rho_\chi$ and $\langle\hat{\chi}^2\rangle$, the field $\zeta$ vanishes ---the noninteracting regime is an example of such regime. Notwithstanding, as the depleted cloud evolves, the condensate response can be decomposed, 
%, which follows a deterministic wave dynamics, is only partially caused by the quantum fluctuations, 
%\col{the rest of it coming from the very condensate fluid nature.} 
%In this sense, t
according to the discussion of 
subsection \ref{qfsec} into a classical and a quantum part, which 
then enables us to identify the {\em part of the force density} %exerted on the system particles 
which comes from quantum fluctuations, as defined in Eq.~\eqref{quantumforce}, 
and which we discuss now. 
\begin{figure}[t]
\includegraphics[scale=0.6]{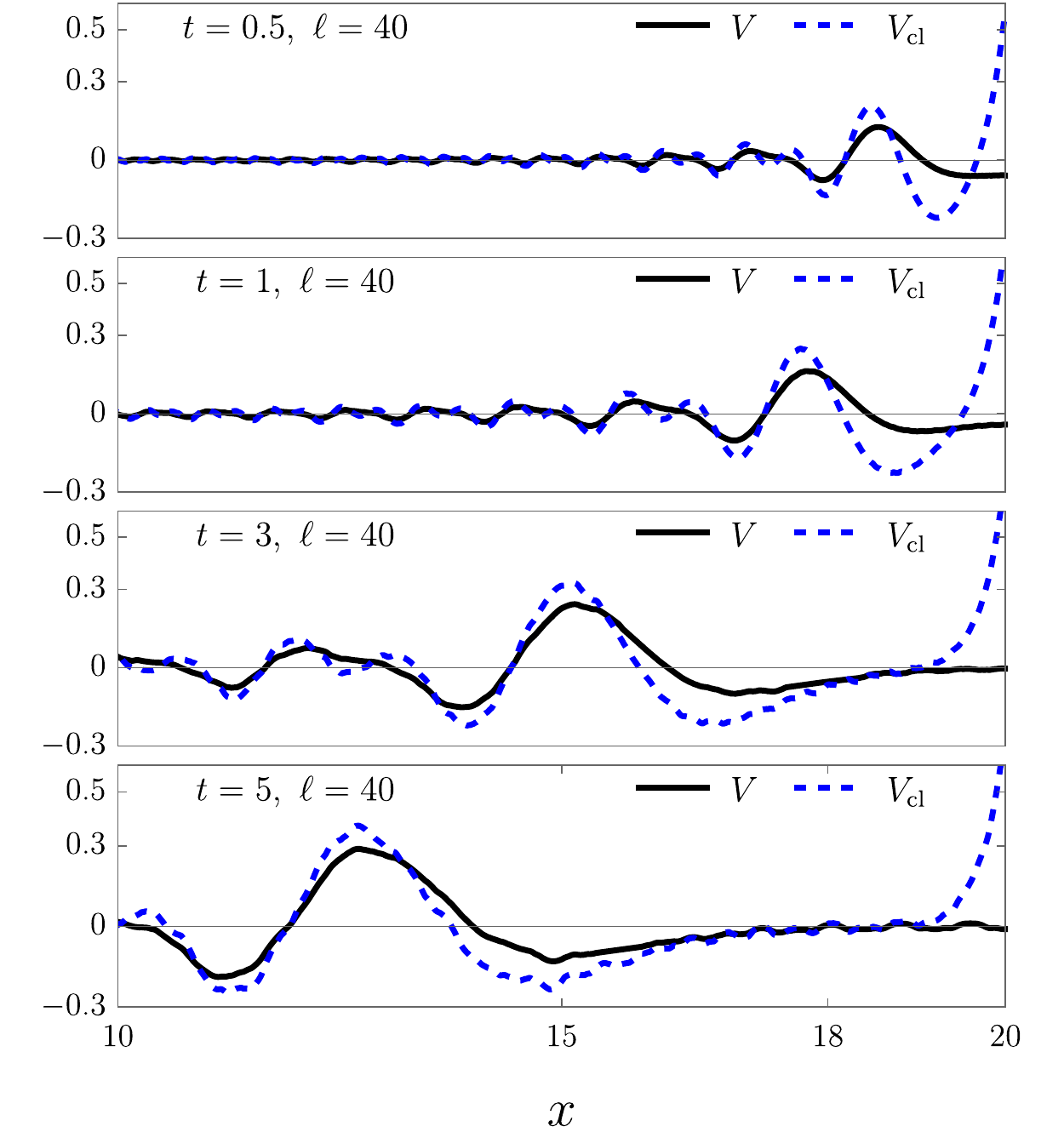}
\caption{Evolution of the total and classical potentials $V$ and $V_{\rm cl}$, respectively, for a condensate of size $\ell=40$ and at several instants of time. The slopes of the curves represent the local force density exerted on the system particles. We note that for the considered time interval, the quantum force has the effect of attenuating the classical Eulerian force, and that this attenuation is more pronounced near the condensate walls.}
\label{figmain6}
\end{figure}  

The total force density is defined by the derivative $\partial_t J=\partial_t J_\zeta$ (given that both $J_0,J_\chi $ vanish), and we notice from Eq.~\eqref{jzetaex} a mathematical issue with  the backreaction analysis: The time derivative of $J_\zeta$ results in a slowly convergent series, preventing the numerical evaluation of the series. In order to circumvent this mathematical difficulty, we can explore the fact that the total force density 
$f_{\rm cl}+f_{\rm q}$ can be expressed as the gradient $\partial_tJ_{\zeta}=-\partial_xV$, where
\begin{equation}
V=\frac{-1}{\ell}\sum_{n=1}^\infty\frac{(-1)^n\cos(2k_nx)}{k_n^2}[\cos(2\omega_n t)-\cos(\omega_{2n} t)],
\end{equation}
is, fortunately, a series with better convergence. Now, it follows from Eq.~\eqref{classical} that for all condensates at rest the property $f_{\rm cl}=f_{\rm cl}(\rho,\partial_x\rho,\partial_x^2\rho,\partial_x^3\rho)$ holds, i.e., the classical force density depends solely on the particle density and its derivatives to our working order $N^{0}$), and here %for our condensate model it 
becomes
\begin{equation}
f_{\rm cl}=-\partial_x\left[\left(1-\frac{\partial_x^2}{4}\right)(\rho_\chi+\rho_\zeta)\right].
\end{equation}
In particular, note that $f_{\rm cl}$ can be determined by (total) density measurements, and 
it is a gradient, $f_{\rm cl}=-\partial_xV_{\rm cl}$, where 
\begin{widetext}
\begin{equation}
V_{\rm cl}=\frac{1}{\ell}\sum_{n=1}^\infty\frac{(-1)^n\cos(2k_nx)}{k_n^2}\left\{\frac{\omega_{2n}^2}{4\omega_n^2}[1-\cos(2\omega_n t)]-1+\cos(\omega_{2n} t)\right\}.
\end{equation}
\end{widetext}
Hence, $V$ represents the actual potential for the system particles, whereas $V_{\rm cl}$ is the expected potential if the particles were solely subjected to an Eulerian dynamics, therefore enabling the study of quantum force effects by comparing $V$ and $V_{\rm cl}$, as presented in Fig.~\ref{figmain6} for a condensate of size $\ell=40$.

We recall that the slopes in the plots of $V$ and $V_{\rm cl}$ represent the total and classical force densities, respectively. Hence, from Fig.~\ref{figmain6} we observe that the classical force is the dominant contribution on the condensate, and the overall effect of the quantum force is the attenuation of $f_{\rm cl}$. In fact, this attenuation is even stronger near the condensate wall, where the observed force vanishes (no flux of particles) and the classical force is stronger, showing that near boundaries 
%\col{which kind of "inhomogeneities} 
the quantum backreaction effects are more pronounced.

Concluding, because both $J_{\zeta}$ and the classical force are determined by the total density $\rho$, we  anticipate the possibility of measuring the quantum force from density measurements alone via the route presented in the above.   
%f_{\rm q}=-\frac{6}{\ell}\sum_{n=1}^\infty\frac{(-1)^nk_n\sin(2k_nx)}{k_n^2+4}[1-\cos(2\omega_n t)]
%\end{equation}

%\begin{equation}
%J_{\rm q}=\frac{-6}{\ell}\sum_{n=1}^\infty\frac{(-1)^nk_n\sin(2k_nx)}{k_n^2+4}\left[t-\frac{\sin(2\omega_n t)}{2\omega_n}\right]
%\end{equation}

\section{Conclusion and Final Remarks}
\label{remarks}

%In this work 
%we extended the backreaction scheme presented in \cite{Fischer2005} to quasi-1D condensates. We 
We have shown that a {\em quantum force density}, signalling a departure from Eulerian 
classical hydrodynamics due to quantum fluctuations, can be written solely in terms of quantities that have direct interpretation: The quantum depletion, the phonon flux, and the density fluctuations. Furthermore, we presented a discussion regarding the proper construction of the Cauchy problem for the backreaction analysis, and its relation with condensate preparation at an experimental level. Our model was then applied to a finite size uniform density condensate, which represents an exactly soluble model and allows for a straightforward interpretation of the results.

As regards backreaction in BECs, the phenomenon of condensate phase diffusion plays a prominent role. Indeed, the existence of finite size interacting quasi-1D condensates is associated with the continuous degradation of the system off-diagonal-long-range-order, which in particular implies that these quantum gases are not stationary systems. Accordingly, because the condensate spontaneously degrades, one has to add the initial configuration of the gas to the backreaction equations to study the system evolution. In order to circumvent this mathematical nuance, we assumed that in our solved model the condensate was initially in a non-interacting regime, for which the system is indeed stationary and can be taken
to reside in its ground state. By starting from this well-defined initial configuration, the required interacting regime can be accessed by driving the system out of equilibrium.

Among the consequences of the solutions of our model, we quote in particular that it highlights the 
problem of distinguishing condensate corrections from quantum depletion in measurements accessing
properties of the {\em total} density (such as its power spectrum), as shown in subsection \ref{gasdensity}. Furthermore, even though our uniform density condensate was initially at rest, backreaction from the quantum fluctuations gives rise to a condensate current dictated by number-conservation, from which the total force density on the system was determined as a gradient of a potential function. Also, by explicitly simulating the Eulerian force corresponding to the observed particle density $\rho=\rho_0+\rho_\chi+\rho_\zeta$, it was possible to conclude that the quantum force density on the system particles attenuates the classical force, an effect more pronounced near the condensate walls.        

An immediate application of the present approach is furnished by considering 
its consequences in analogue gravity say for the backreaction of emitted Hawking radiation 
onto the condensate background \cite{Balbinot,Tricella}. 
We also note in this connection that {\em classical} backreaction has already been studied experimentally in shallow water tanks, cf.~in Ref.~\cite{Patrick}.

 %Nevertheless, we note that the ground state definition within the number-conserving expansion we adopted has certain subtleties linked to the system initial condition. Indeed, by inserting the field expansion $\hat{\Psi}=\phi_0+\hat{\chi}+\zeta$ in the system Hamiltonian (which can be read from \cite{Castin1998}, for instance), we obtain the ground state energy
%
%\begin{widetext}
%\begin{align}
%E_0=\int\d x\left\{\left[\frac{|\partial_x\phi_0|^2}{2}+\left(U+\frac{g\rho_0}{2}\right)|\phi_0|^2\right]+\frac{i}{2}\langle\hat{\chi}^\dagger\partial_t\hat{\chi}-(\partial_t\hat{\chi}^\dagger)\hat{\chi}\rangle+i(\zeta^*\partial_t\phi_0-\zeta\partial_t\phi_0^*)\right\}+\mathcal{O}(N^{-1/2}),
%\end{align}
%\end{widetext}
%
%that depends explicitly on $\zeta$ up to our working order $N^{0}$.

\section{Acknowledgments} 
This work has been supported by the National Research Foundation of Korea under 
Grants No.~2017R1A2A2A05001422 and No.~2020R1A2C2008103.

\appendix

\section{Canonical commutation relation}
\label{completeness}

In this Appendix we show that the set $\{\Phi_n\}_n$ of field modes constructed in the SubSec.~\ref{fieldmodes} is complete, namely, the corresponding quantum field expansion satisfies Eq.~\eqref{ccr} for $-\ell/2<x,x'<\ell/2$. To this end, it is sufficient to show that $[\hat{\psi}(t,x),\hat{\psi}^\dagger(t,x')]=\delta(x-x')$ holds. By means of Eq.~\eqref{qfield}, we have
\begin{align}
[\hat{\psi}(t,x)&,\hat{\psi}^\dagger(t,x')]=\nonumber\\
&\sum_{n=0}^\infty\left[u_n(t,x)u^*_n(t,x')-v^*_n(t,x)v_n(t,x')\right].
\end{align}
Let us consider first the noninteracting regime, $t<0$. In this regime, the field modes are given by Eq.~\eqref{modenon}, and we find
\begin{align}
[\hat{\psi}(t,x)&,\hat{\psi}^\dagger(t,x')]=\nonumber\\
&\frac{1}{2\ell}\sum_{n=-\infty}^\infty\left[e^{in\pi\Delta x/\ell}+(-1)^ne^{in\pi(x+x')/\ell}\right].\label{aux}
\end{align}
We can write the equation above in terms of delta functions using Poisson's summation formula \cite{Lorenci}
\begin{equation}
\frac{1}{2\ell}\sum_{n=-\infty}^\infty e^{in\pi y/\ell}=\sum_{n=-\infty}^\infty\delta(y-2\ell n),
\end{equation}
from which we obtain
\begin{align}
[\hat{\psi}&(t,x),\hat{\psi}^\dagger(t,x')]=\nonumber\\
&\sum_{n=-\infty}^\infty\left[\delta(\Delta x-2\ell n)+\delta(x+x'-2\ell n-\ell)\right].
\end{align}
By inspecting the R.H.S. of the equation above we conclude that for $-\ell/2<x,x'<\ell/2$ all the delta functions give zero contribution except $\delta(\Delta x)$, thus verifying that the field is canonically quantized in the noninteracting regime. 

In the interacting regime, the field modes assume the form in Eqs.~\eqref{series} and \eqref{coeffm}, and straightforward manipulations lead for $[\hat{\psi}(t,x),\hat{\psi}^\dagger(t,x')]$ to the identical Eq.~\eqref{aux}.

\section{Building the condensate corrections}
\label{cr}

In this Appendix we show how to derive the main result of our work, the solution of Eq.~\eqref{GPcorr} in the interacting regime, presented in Eq.~\eqref{zetasol}. 
We start by making the ansatz 
\begin{equation}
\zeta(t,x)=\exp(-i\mu t)f(t,x)/\sqrt{\rho_0},
\end{equation} 
where $f$ is a solution of
\begin{equation}
i\partial_tf=-\frac{1}{2}\partial_x^2f+(f+f^*)+2\rho_\chi+
\langle\hat{\psi}^2\rangle,
\end{equation}
with $f=0$ at $t=0$, and $\partial_xf=0$ at $x=\pm\ell/2$ for all times. Inspired by how the solutions of Eq.~\eqref{BdGred} pertaining to the BdG equation were built, we define the spinor $F=(f,f^*)^{\rm t}$, which is a solution of
\begin{equation}
i\partial_t\sigma_3F=-\frac{1}{2}\partial_x^2F+\sigma_4F+
\left(
\begin{array}{c}
2\rho_\chi+\langle\hat{\psi}^2\rangle\\
2\rho_\chi+\langle\hat{\psi}^{\dagger2}\rangle
\end{array}\right).\label{tosolve}
\end{equation}
Thus our goal is to solve Eq.~\eqref{tosolve} subject to $F=0$ at $t=0$ and with the Neumann boundary conditions imposed on $F$.

It follows from the quantum field expansion of Eq.~\eqref{qfield} for $t>0$ that
\begin{widetext}
\begin{equation}
2\rho_\chi+\langle\hat{\psi}^2\rangle=\frac{t^2-it}{\ell}+\frac{1}{\ell}\sum_{n=1}^\infty\frac{(-1)^n}{k_n^2(k_n^2+4)}\left[(-1)^n+\cos(2k_n x)\right]\left\{(2-k_n^2)[1-\cos(2\omega_n t)]-2i\omega_n\sin(2\omega_nt)\right\}.\label{part}
\end{equation}
\end{widetext}
Notice that from Eq.~\eqref{part} the quantity $2\rho_\chi+\langle\hat{\psi}^2\rangle$ is a sum over the various field mode contributions, and because Eq.~\eqref{tosolve} is linear, $F$ can be constructed by summing the solutions of Eq.~\eqref{tosolve} for each term in Eq.~\eqref{part} (indexed by $n$) subjected to the initial and spatial boundary conditions. Accordingly, we write $F(t,x)=\sum_{n=0}^\infty F_n(t,x)$, and solve for each $F_n$ such that $F_n(0,x)=0$ and $F_n$ satisfies Neumann boundary conditions at $x=\pm\ell/2$.

We now determine each $F_n$. For $n=0$ the result is straightforwardly obtained, 
%and we simply note that 
$F_0=-(t^2/2\ell)(1,1)^{\rm t}$. 
For $n>0$ we need to solve
\begin{widetext}
\begin{align}
&\left(i\partial_t\sigma_3+\frac{1}{2}\partial_x^2-\sigma_4\right)F_n\nonumber\\
&=\frac{(-1)^n[(-1)^n+\cos(2k_n x)]}{\ell k_n^2(k_n^2+4)}\left[
(2-k_n^2)\left(
\begin{array}{c}
1\\
1
\end{array}\right)
-e^{2i\omega_n t}\left(
\begin{array}{c}
\omega_n-k_n^2/2+1\\
-\omega_n-k_n^2/2+1
\end{array}\right)
-e^{-2i\omega_n t}\left(
\begin{array}{c}
-\omega_n-k_n^2/2+1\\
\omega_n-k_n^2/2+1
\end{array}\right)
\right].\label{eqfn}
\end{align}
\end{widetext}
Suppose that $\tilde{F}_n$ is a given particular solution for the equation above. Then the required function $F_n$ must be of the form
\begin{equation}
F_n=a_0\Pi_0+b_0\tilde{\Pi}_0+\sum_{j=1}^\infty[a_j\Pi_j-b_j\sigma_1\Pi^*_j]+\tilde{F}_n,
\label{FnExpansion}
\end{equation} 
using that the set of field modes $\{\Pi_0,\tilde{\Pi}_{0},\Pi_n,\sigma_1\Pi_n^*\}$ is complete as discussed in Sec.~\ref{fieldmodes}. 
The coefficients $a_j, b_j$ are uniquely defined by the initial condition $F_n(0,x)=0$, which leads to a Fourier expansion of $-\tilde{F}_n(0,x)$, in the exactly same fashion as presented in Sec.~\ref{fieldmodes}, namely,
\begin{equation}
a_0\Pi_0+b_0\tilde{\Pi}_0+\sum_{j=1}^\infty[a_j\Pi_j-b_j\sigma_1\Pi^*_j]=-\tilde{F}_n
\end{equation}
%
%\pagebreak
where
\begin{equation}
\begin{aligned}
&a_{0}=-\frac{\langle\tilde{\Pi}_0,\tilde{F}_n\rangle}{\langle\tilde{\Pi}_0,\Pi_0\rangle},\ \ \ a_{j}=-\langle\Pi_j,\tilde{F}_n\rangle,\\
&b_{0}=-\frac{\langle\Pi_0,\tilde{F}_n\rangle}{\langle\Pi_0,\tilde{\Pi}_0\rangle},\ \ \ b_{j}=-\langle\sigma_1\Pi_j^*,\tilde{F}_n\rangle, \label{Coeff}
\end{aligned}
\end{equation}
and all the functions are evaluated at $t=0$. Moreover, because $\tilde{F}_n=\sigma_1\tilde{F}_n^*$, we should have $b_{j}=-\langle\sigma_1\Pi_j^*,\tilde{F}_n\rangle=b_{j}=\langle\Pi_j^*,\sigma_1\tilde{F}_n\rangle=\langle\Pi_j,\sigma_1\tilde{F}^*_n\rangle^*=\langle\Pi_j,\tilde{F}_n\rangle^*=-a_j^*$. 

Therefore the problem is reduced to the determination of the particular solution $\tilde{F}_n$, which can be obtained as follows. We first note that the R.H.S. of Eq.~\eqref{eqfn} is the sum of constant terms that depend on $x$ or $t$ only, and terms that depend on both $x$ and $t$. Thus one strategy for finding a $\tilde{F}_n$ is to use the field equation's linearity and solve for each term separately, which can be done in a straightforward manner. We find that
\begin{widetext}
\begin{align}
\tilde{F}_n=-\frac{(-1)^n}{2\ell k_n^2(k_n^2+4)}\bigg\{&(-1)^n\left[
(2-k_n^2)\left(
\begin{array}{c}
1\\
1
\end{array}\right)
-\frac{e^{2i\omega_n t}}{\omega_n}\left(
\begin{array}{c}
\omega_n-k_n^2/2\\
\omega_n+k_n^2/2
\end{array}\right)
-\frac{e^{-2i\omega_n t}}{\omega_n}\left(
\begin{array}{c}
\omega_n+k_n^2/2\\
\omega_n-k_n^2/2
\end{array}\right)
\right]\nonumber\\
%%%%%%%%%%%%%%%%%%%%%
&+2\cos(2k_n x)\left[
\frac{2-k_n^2}{1+k_n^2}\left(
\begin{array}{c}
1\\
1
\end{array}\right)
+\frac{e^{2i\omega_n t}}{k_n^2}\left(
\begin{array}{c}
-\omega_n+k_n^2/2\\
\omega_n+k_n^2/2
\end{array}\right)
+\frac{e^{-2i\omega_n t}}{k_n^2}\left(
\begin{array}{c}
\omega_n+k_n^2/2\\
-\omega_n+k_n^2/2
\end{array}\right)
\right]\bigg\}\label{psol}
\end{align}
\end{widetext}
is a particular solution to Eq.~\eqref{eqfn}. 
%Finally, by u

Using this particular solution and the modes given by Eqs.~\eqref{Pi0}, \eqref{pdmode}, and \eqref{Pin}, 
we find the set of coefficients \eqref{Coeff} in the expansion \eqref{FnExpansion}: 
%\pagebreak
%
\begin{align}
& a_0=0,\qquad 
b_0=-\frac{1}{\ell (k_n^2+4)}, \\
%\end{equation}
%and
%
%\begin{equation}
& a_j=-\delta_{j,2n}\frac{2(-1)^n(2-\omega_{2n}+k_{2n}^2/2)}{\sqrt{2\ell[1-(\omega_{2n}-k_{2n}^2/2-1)^2]}\omega_{2n}^2},
\end{align}
which give rise to the solution presented in Eq.~\eqref{zetasol} of the main text.

\bibliography{bbQ8}

%%%%%%%%%%%%%%%%%%%%%%%%%%%%%%%%%%%%%%%%%%%%%%%%%%%%%%%%%%%%%%%%%%%%%%%%%%%%%%%%%%%%%%%%%%%%%%%%%%%%%%%%%%%%%%%%%%%%%%%%%%%%%%%%%%
%
%%%%%%%%%%%%%%%%%%%%%%%%%%%%%%%%%%%%%%%%%%%%%%%%%%%%%%%%%%%%%%%%%%%%%%%%%%%%%%%%%%%%%%%%%%%%%%%%%%%%%%%%%%%%%%%%%%%%%%%%%%%%%%%%%%
%
%%%%%%%%%%%%%%%%%%%%%%%%%%%%%%%%%%%%%%%%%%%%%%%%%%%%%%%%%%%%%%%%%%%%%%%%%%%%%%%%%%%%%%%%%%%%%%%%%%%%%%%%%%%%%%%%%%%%%%%%%%%%%%%%%%

%\end{document}
%\bibliography{prls}
%\vspace*{20em} 
%\newpage 
%\appendix
%%%%%\vspace*{100em} 
%%%%%\newpage
%%%%%\begin{widetext}
%%%%%\setcounter{equation}{0}
%%%%%\setcounter{figure}{0}
%%%%%\setcounter{table}{0}
%%%%%\setcounter{page}{1}
%%%%%\renewcommand{\theequation}{S\arabic{equation}}
%%%%%\renewcommand{\thefigure}{S\arabic{figure}}

\end{document}